\documentclass[11pt]{article}

\usepackage[a4paper,hmarginratio=1:1,vmarginratio=2:3,totalwidth=16cm,
totalheight=22.cm]{geometry}

\usepackage{latexsym}
\renewcommand{\baselinestretch}{1.2}   
\usepackage{epsfig}

\newcommand{\cI}{{\cal I}}

\newcommand{\cL}{{\cal L}}
\newcommand{\cM}{{\cal M}}
\newcommand{\cN}{{\cal N}}
\newcommand{\cO}{{\cal O}}
\newcommand{\cP}{{\cal P}}

\newcommand{\cV}{{\cal V}}
\newcommand{\cZ}{{\cal Z}}

\newcommand{\Tr}{\mbox{Tr}}

\newcommand{\be}{\begin{equation}}
\newcommand{\ee}{\end{equation}}
\newcommand{\bea}{\begin{eqnarray}}
\newcommand{\eea}{\end{eqnarray}}

\newcommand{\ra}{\rightarrow}

\newcounter{oldcounter} 
\addtocounter{equation}{1}
\begin{document} 
\begin{flushright}  
{OUTP-0603P, CERN-PH-TH/2006-162}\\
\end{flushright}  
\thispagestyle{empty}

\vspace{3cm}
\begin{center}
{\Large {\bf Living with Ghosts and their Radiative   Corrections}}\\
\vspace{1.cm}

{\bf I. Antoniadis$^{\,a, b}$, E. Dudas$^{\,a, b, c}$,  D.~M. Ghilencea$^{\,d}$}\\
\vspace{0.5cm}
{\it $^a $Department of Physics, CERN - Theory Division, 1211 Geneva 23, Switzerland.}\\[6pt]
{\it $^b $CPHT, UMR du CNRS 7644,  \'Ecole Politechnique, 91128
  Palaiseau Cedex, France.}\\[6pt]
{\it $^c $LPT, UMR du CNRS 8627, B\^at 210, Universit\'e de Paris-Sud, 
91405 Orsay Cedex, France} 
\\[6pt]
{\it $^d $Rudolf Peierls Centre for Theoretical Physics, University of
  Oxford,\\
1 Keble Road, Oxford OX1 3NP, United Kingdom.}\\

\end{center}
\vspace{0.5cm}
\begin{abstract}
\noindent
The  role of higher derivative operators in 4D effective field 
 theories is discussed in both  non-supersymmetric and supersymmetric
 contexts. The approach, formulated in the Minkowski space-time,
shows that theories with higher derivative operators do not always have 
an improved UV behaviour, due to subtleties related to the analytical continuation
from the Minkowski to the Euclidean metric. This continuation
 is further affected at the dynamical level due to a field-dependence of
 the poles of the  Green functions of the particle-like states,
for curvatures of the  potential of order unity in ghost mass units.
 The one-loop scalar potential in $\lambda \phi^4$ theory with a 
single  higher derivative  term  is shown to have infinitely many
 counterterms, while for a very large  mass of the ghost the usual
4D renormalisation is recovered.
In the supersymmetric context of the O'Raifeartaigh model of spontaneous
supersymmetry breaking with a higher derivative (supersymmetric)
 operator,  it is found  that quadratic divergences are present in the
 one-loop self-energy of the scalar field. They arise with a
 coefficient proportional to  the amount of supersymmetry
 breaking  and suppressed by the scale of the higher
 derivative operator. This is also true in the
 Wess-Zumino model with higher derivatives and explicit soft breaking
of supersymmetry.  In both models, the  UV logarithmic behaviour  is 
restored in the
 decoupling limit of the ghost.
\end{abstract}
\newpage

\setcounter{page}{1}

\section{Introduction}

Low energy supersymmetry can provide  a solution to the hierarchy
problem, when this is softly broken at around the TeV scale, and future 
LHC experiments will be able to test some of the predictions 
associated with this. In general, low energy models such as 
Standard Model or its supersymmetric versions, 
are regarded as  effective theories valid up to some high
mass scale,  at which footprints of a more fundamental theory can show up.
These can be in the form of higher dimensional operators, 
and they  can play an important role in clarifying the more
fundamental theory valid beyond the high scale, which suppresses their
effects at low energies.
Usually only {\it non-derivative} higher dimensional operators are
considered in 4D,  while higher derivative ones are less studied or
popular,  due to many difficult issues involved:
 either (high scale) unitarity or causality violation, non-locality, 
the role of the additional ghost fields  present in the context of field 
theories with higher derivative operators (see for example
\cite{swh}-\cite{KKP}), etc.
Perhaps the most difficult issue  in such  theories is 
related to  the analytical continuation from the Euclidean to
the Minkowski space (or vice-versa) which are not necessarily in a
one-to-one correspondence.  In this work we attempt to address some 
of these problems.

Another motivation for studying higher dimension derivative operators
in the 4D action comes from the compactification of higher dimensional
gauge theories.
If physics at low scales  is the low-energy limit of a more
fundamental such theory, higher derivative operators can be 
generated  dynamically,  even in the simplest (orbifold)
compactifications. 
For example, higher derivative operators are generated
 by gauge (bulk) interactions in 6D at one-loop or  5D beyond one-loop
\cite{nibbelink,Ghilencea:2006qm,santamaria,Alvarez:2006we}. 
Brane-localised higher derivative operators are also 
generated at the loop level \cite{Ghilencea:2004sq}, by 
superpotential interactions  in 5D or 6D orbifolds. Higher derivative
interactions were also studied in the context of   Randall-Sundrum
models \cite{RS}.
Therefore, clarifying the role of such operators can help a better
understanding of compactification.
Further, higher derivative operators are important in other
studies:  cosmology
\cite{Anisimov:2005ne,Gibbons:2003yj}, phase transitions and Higgs mechanism
\cite{BMP,c1,c5}, supergravity/higher derivative gravity 
 \cite{Stelle}-\cite{Avramidi:1985ki}, string theory 
\cite{Eliezer:1989cr,Polyakov:1986cs},  used as regularisation
method \cite{Slavnov:1977zf}, etc. We therefore consider
it is worth  investigating the role of higher derivative operators in a  4D
effective field  theory, be it supersymmetric or not.

One common problem  in theories with higher derivative
operators is that they are in many
 cases  formulated and studied in the Euclidean space,
and then it is {\it assumed} that there exists an analytical continuation to the
Minkowski space. In some simple cases the results of such studies might hold true upon
analytical continuation to the Minkowski space.
However, in the presence of higher derivative
operators, the dispersion relations change, new poles in the
fields' propagators are present, and the position of some of these 
can move in the complex plane and become field dependent.
In this situation, the analytical continuation to Minkowski
space becomes  more complicated and possibly
ambiguous, and one can be faced with  difficult choices: different
$\pm i\,\epsilon$ prescriptions in  the Green
functions can lead to different results upon continuation  from the Euclidean
to  the Minkowski space. To avoid this, we start instead with a formulation in 
the Minkowski space, and pay 
special attention to this problem. We do so by making the simple
observation that the partition function for the second order theory 
in the {\it Minkowski} space, should  be well-defined
and also  recovered from 
the fourth order Minkowski  action, in the decoupling limit of the higher
derivative terms,  when the scale suppressing these operators  
is taken to infinity. This  is consistent with the finding  \cite{Hawking:2001yt}
that a fourth order theory  can make sense in 
Minkowski space if one treats it like a second order theory.
 Ensuring that the model {\it without} higher
derivatives (or with their degrees of freedom integrated out)
has a well defined partition function in the Minkowski
space turns out to be enough to fix the ambiguities mentioned
earlier. No special 
assumptions regarding analytical continuation are made, for this is 
fixed unambiguously, with some  interesting results.

The plan of the paper is the following:
we first consider a 4D scalar field theory with a higher
derivative term in the action and  study its Euclidean and Minkowski 
formulations and  their relationship.
Section~\ref{one-loop-potential} addresses the one-loop effects of
 higher derivative terms on the scalar
field self-energy  and scalar potential. We considered interesting  to
extend the   analysis to the supersymmetric case (Section~\ref{O'R})
 and investigate the role of supersymmetric 
 higher derivative terms in the case of
 O'Raifeartaigh model of spontaneous supersymmetry breaking. We  analyse the 
 role of these operators for the UV behaviour of the self-energy of
 the scalar field.  Section~\ref{WZ}  addresses the same problem in the
 case of a Wess-Zumino model with supersymmetric higher derivatives
terms and explicit soft supersymmetry-breaking terms. In both models it was
found that, despite the soft nature of supersymmetry breaking,
UV quadratic divergences are generated for the scalar field self-energy,
with a coefficient related to the amount of supersymmetry
breaking and suppressed by the (high) scale of the higher derivative operator.

\section{Effects of Higher Derivative Operators at the loop level.}
\subsection{A simple example: Higher derivative terms  in scalar field
  theory.}\label{non-susy}

Let us start with
 a simple 4D toy model with a higher derivative term in the
$\lambda\phi^4$  theory and study its effects at the classical and 
quantum level. Our starting point is  the 4D Lagrangian
\smallskip
\begin{eqnarray}\label{action_4}
\cL= -\frac{1}{2}\, \phi \,(\xi\,\Box^2+\Box)\,\phi - V(\phi),
\qquad\quad 
V(\phi)\equiv V_0 + \frac{1}{2} \,(m^2 - i\,\epsilon) \, \phi^2 
+\frac{\lambda}{4!}\,\phi^4 
\end{eqnarray}

\medskip
\noindent
with $\xi\equiv 1/M^2_*>0$  and $M_*$ is some high scale of 
``new physics'' where the higher derivative
term becomes  important. Our metric convention is
$(+,-,-,-)$ and $\Box\equiv\partial_\mu \partial^\mu$.
Additional, higher order derivative terms can be added, but these are
 suppressed by higher powers of the scale $M_*$. In the  limit
$\xi\rightarrow 0$, the higher derivative terms decouple at the 
classical level.

The term $+i\,\epsilon \,\phi^2$ in $\cL$ (we take $\epsilon>0$)
is consistent with the requirement
 that the partition function for the particle state $\phi$ of the 
second-order theory (i.e. in the limit $\xi\!=\!0$),
as defined in the   Minkowski space-time $Z\sim \int D(\phi) \exp(i \int
d^4 x \,\cL)$, remains convergent at {\it all} momentum 
scales\footnote{This is nothing but the usual
  prescription $+i\,\epsilon$  in the scalar propagator
 of 2$^{nd}$ order theory (Minkowski space).}.
The presence of $+i \epsilon\phi^2$ is not a choice, it 
is  just the familiar  prescription  present in $\cZ$ in second order
theories. It is then natural to require that this prescription
  remain valid in our case too (i.e. for non-zero $\xi$) and
at all scales,  and this is the only
   assumption\footnote{This assumption  is consistent with  treating 
the partition function $\cZ$ of a 4$^{th}\!$ order theory in {\it
  Minkowski} space as
 one of a  2$^{nd}\!$ order theory  \cite{Hawking:2001yt} of $\xi=0$,
 with the ghost state $\Box\phi$ as a virtual state  rather than
an asymptotic one.} we make.
This observation  is important, for it
fixes  potential ambiguities  present in  some treatments of 
theories with higher order derivatives.
Further,  in (\ref{action_4})  one can change the basis
\cite{Hawking:2001yt} to $\varphi_{1,2}$
\medskip
\begin{eqnarray}\label{transition}
\varphi_{1,2}=-\frac{(\Box +m_{\pm}^2\pm i\,\epsilon^*(0)) \,\phi \,\sqrt \xi}{
  (m_+^2-m_-^2+2 i \epsilon^*(0))^{1/2}},
\end{eqnarray}
where we introduced
\begin{eqnarray}\label{mpm}
m_{\pm}^2
\equiv\frac{1}{2 \xi}\Big[1 \pm
  (1-4\,\xi\,m^2)^{\frac{1}{2}}\Big]>0,\qquad
\epsilon^*(0)\equiv\frac{\epsilon}{(1-4 \xi m^2)^{1/2}}>0
\end{eqnarray}
Then eq.(\ref{action_4})  becomes
\begin{eqnarray}\label{action_4g}
\cL=-\frac{1}{2} \varphi_1 (\Box +m_-^2 -i \epsilon^*(0)) \,
\varphi_1+\frac{1}{2} \varphi_2
(\Box+m_+^2+i\,\epsilon^*(0))\,\varphi_2 -\bigg[V_0+\frac{\lambda}{4!} \,\,
  \frac{(\varphi_1-\varphi_2)^4}{1-4\,\xi\,m^2}\bigg]
\end{eqnarray}

\medskip
\noindent
Therefore $\varphi_2$ is a ghost, it has the ``wrong'' sign in front of
the kinetic term. 
Note the presence in the interaction  of a coupling 
between $\varphi_{1}$ and $\varphi_2$,  which prevents one from ignoring 
the effects of $\varphi_2$. $\varphi_{1,2}$ are not
independent  degrees of freedom, since $\phi$ and $\Box\phi$ entering 
their definition are not. In fact only $\phi\propto \varphi_2-\varphi_1$
exists as an asymptotic
state \cite{Hawking:2001yt}.  To conclude, our original field
$\phi$ is  a ``mixing'' of particle-like ($\varphi_1$) 
 and ghost-like ($\varphi_2$) states, eq.(\ref{transition}).

 In basis  (\ref{action_4g}) the ghost' presence  is
manifest, but for   technical reasons it is easier to work in basis
(\ref{action_4})  where the presence of this degree of freedom is implicit
in the propagator of $\phi$ alone, defined by (\ref{action_4}).
Note  the emergence from (\ref{action_4})  of $\pm i \epsilon^*(0)$ terms
  in (\ref{action_4g}), important later on. 
These are  usually  overlooked in the literature: if one started
instead with  action (\ref{action_4g})
 without these prescriptions,  one could face 
a {\it choice} ($\pm$)  for the  prescription in the ghost propagator.
In our case these prescriptions are {\it derived}  from the 
4$^{th}$ order action of (\ref{action_4})  which is our starting theory.
Further, for fixed $\epsilon\ll 1$, the condition
 $\epsilon^*(0)\ll 1$  requires $m^2\xi\ll 1/4$, which we assume
to hold true. 
For model building one would prefer that effects associated 
with the mass of the ghost  (unitarity violation, etc) be suppressed 
by a high scale, therefore we take $m_+^2\sim 1/\xi\gg m_-^2\sim m^2$.

In principle  one could start with action (\ref{action_4g}) and
  insist to have  a $-i\epsilon^*(0)$  
prescription in its ghost term and study such theory; however, with
(\ref{transition}) changed accordingly to reflect this,
the relation  of such theory  to (\ref{action_4}) is changed:
one would obtain in (\ref{action_4}) a {\it momentum dependent}
pole prescription, of type
$\epsilon\!\ra\!\epsilon'\!=\!\epsilon^*(0) (1+\xi \Box)$, unlike in our
 starting  theory (\ref{action_4})
 where $\epsilon$ is momentum  independent.
 At $p^2\!\ll\! 1/\xi$ such  theory
would be  similar to (\ref{action_4g}) (ghost being decoupled), but
at $p^2\!\gg \! 1/\xi$, its partition function for $\phi$ 
would not remain convergent in Minkowski space, without
 further assumptions. Such theory can be  interesting
but will not be discussed in this work.

Let us calculate   the one-loop  correction to 
the mass of $\phi$ using  the   ``basis'' of eq.(\ref{action_4})
\smallskip
\begin{eqnarray}\label{delta-mass-M}
 -i\, \delta m^2& =& 
-i \lambda\, \mu^{4-d}\int \frac{d^d p}{(2\pi)^d}
\frac{i}{-\xi p^4 +p^2-m^2+i\epsilon}\nonumber\\[9pt]
&=&\frac{-i \lambda\,\mu^{4-d}}{\sqrt{1-4 \xi m^2}}
 \int  \frac{d^d p}{(2\pi)^{d}}
\bigg[
\frac{i}{p^2-m_-^2+i \epsilon^*(0)}-\frac{i}{p^2-m_+^2-i \epsilon^*(0)}
\bigg]\nonumber\\[9pt]
&=&
\frac{-i \lambda\,\mu^{4-d}}{\sqrt{1-4 \xi m^2}}\int_{\bf E}
\frac{d^d p}{(2\pi)^d} \bigg[\frac{1}{p^2+m_-^2}+\frac{1}{p^2+m_+^2}\bigg],
\end{eqnarray}

\smallskip
\noindent
where the last integral is in the Euclidean space, showed
 by the index {\bf E}, while $\mu$ is the standard non-zero
finite mass scale introduced by the DR scheme with $d=4-\omega$,
 $\omega\ra 0$.
 The above result written instead  in a  cutoff regularisation
makes explicit the nature of (scale) divergence:
\medskip
\begin{eqnarray}\label{delta-mass-E}
-i \,\delta m^2
&=&\frac{-i \lambda}{(4\pi)^2 \sqrt{1-4\, \xi \, m^2}}
\bigg[\Big[\Lambda^2-m_-^2\ln\Big(1+\Lambda^2/m_-^2\Big)\Big]
+\Big[\Lambda^2-m_+^2\ln\Big(1+\Lambda^2/m_+^2\Big)\Big]\bigg].
\end{eqnarray}

\medskip
\noindent
In the formal limit $\xi \rightarrow 0$ i.e. $m_+^2\gg \Lambda^2$, the 
last two terms (i.e. the ghost correction) cancel against each 
other\footnote{This is easily seen by a Taylor expansion of the 
logarithm of argument close to unity.},  to leave only the particle
part, since then $m_-^2\rightarrow m^2$. Note that in any other case,
 the ghost contributes to the quadratic divergence of the correction. 
This is important and  contradicts the common statement  
that theories with higher derivative operators have an improved UV regime, 
as naively expected from power-counting in Minkowski space, 
first line in eq.(\ref{delta-mass-M}): this eq would  instead suggest a $\ln\Lambda$ 
divergence only! The explanation of this difference is 
the plus sign in the last  eq above, due to a Wick rotation to
 Euclidean space, and the two quadratic divergences add up rather than 
cancel between the two contributions in (\ref{delta-mass-M}), (\ref{delta-mass-E}).

As a separate remark, let us also note that
if we considered the case of 2 dimensions ($d\rightarrow 2$) naive power counting 
applied to first line in (\ref{delta-mass-M}) would suggest the integral is UV finite
 and yet, the integral is
logarithmically divergent in the UV, see the last line in (\ref{delta-mass-M}).
 Therefore, it is not only that diagrams that are already UV divergent
can turn out to have a worse UV behaviour when Wick rotated to Euclidean space,
but also diagrams that appear UV finite  by power counting 
turn out to be UV divergent, with  implications for the renormalisation
algorithm\footnote{The naive application of the power-counting
theorem \cite{Weinberg:1959nj} fails  in the above contexts
since for the theorem to hold, the  Wick rotation
to Euclidean space must yield the Euclidean version of the theory,
at least in the UV.}.

The result  also shows that in this simple case
 the ghost  cannot trigger  negative 
 (mass)$^2$ for the scalar, to bring in an internal symmetry breaking,
 since here $\delta m^2>0$. However this
remains a  possibility  in similar models, if one included 
additional gauge interactions, fermionic contributions of opposite
 sign, or other additional  6D  operators (such as
 for example $\phi^2\Box\phi^2$ with a different coefficient)
when more corrections are present and can trigger  symmetry breaking.

Let us remark that one can also use  a series expansions
of the general propagator  in the presence of the higher derivative term,
and  keep only a leading correction:
\smallskip
\begin{eqnarray}\label{expansion}
\frac{1}{-\xi \,p^4+p^2-m^2+i\epsilon}=
\frac{1}{p^2-m^2+i\epsilon}+\frac{\xi
  \,p^4}{(p^2-m^2+i\,\epsilon)^2}+\cO(\xi^2 p^4)
\end{eqnarray}

\bigskip
\noindent
Here the first term in the rhs is the usual propagator, 
while the effect of the higher derivative contribution is, to the
lowest order,  just a small correction. 
Note however the different number of poles for $p_0$ of the
lhs and rhs in (\ref{expansion}) with implications for continuation
 from the Minkowski  to the Euclidean space. The additional lhs pole,
ultimately brings in an additional degree of freedom (ghost), and is
 present only  upon re-summing all
higher order terms in the rhs.

The  expansion in (\ref{expansion}), when performed under a loop
 integral such as (\ref{delta-mass-M}), is  restrictive for it 
assumes  that one can neglect terms 
$\xi p^2\ll 1$ for {\it any value} of $p^2$,   in particular for
$1/\xi\gg p^2\sim\Lambda^2$ where $\Lambda$ is a UV cutoff of the
integral. 
Using the above expansion in (\ref{delta-mass-M}) one obtains an
approximation (in $\xi$) of the previous result for $\delta m^2$, but
will be valid only for a ghost mass ($\sim 1/\xi$) 
much larger than the cutoff scale.
This situation is unlike the results  of eq.(\ref{delta-mass-M}), 
(\ref{delta-mass-E}), 
which are more general, being valid for any value of the 
mass of the ghost ($m_+^2\sim 1/\xi$)
relative to  $\Lambda$, in particular for
$m_+^2\sim \Lambda^2$, when (\ref{expansion}) is not a good
 approximation
since  it would require many more terms of the series to be included.
To conclude, for applications
one should use  the full propagator in the presence of higher
derivative term and, after that, one can consider  special cases such as the
limit  $m_+^2\sim 1/\xi\gg \Lambda^2$ of decoupling the ghost-like contribution.
The advantage of this approach is that one will
 also be able to consider the case $m_+^2\sim \Lambda^2$
or  $m_+^2< \Lambda^2$, (with $m_+^2$ assumed however  much larger
 than a TeV 
scale, for phenomenological reasons).

Finally, let us now use  the action in  eq.(\ref{action_4g}) and the
basis $\varphi_{1,2}$ (instead of $\phi, \Box\phi$), 
to recover $\delta m^2$ of (\ref{delta-mass-M}).
We thus compute at one-loop $\delta m_{\pm}^2$ given by
interaction  (\ref{action_4g}) proportional to $(\varphi_1-\varphi_2)^4$.
  There are two one-loop  diagrams of order $\lambda$ which
contribute to $\delta m_-^2$:
one has in the loop a propagator of $\varphi_1$, and a
symmetry factor 12; the second Feynman 
diagram has $\varphi_2$ in the loop, with a symmetry
factor of 2. 
\begin{figure}[t]{
\centerline{\psfig{figure=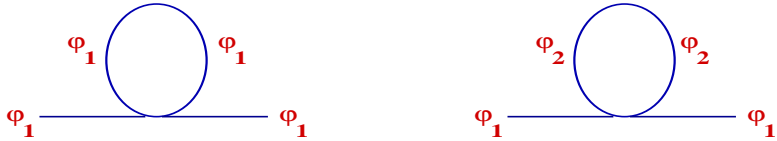,height=0.7in,width=4.in,angle=0}}
\def\baselinestretch{1.1}
\caption{\small The  Feynman diagrams which contribute 
to the self-energy $\delta m_-^2$ of 
 the particle-like field $\varphi_1$.} 
\label{fig00}}
\end{figure}
The result for $\delta m_-^2$ is given below:
\medskip
\begin{eqnarray}
- \frac{i}{2} \,\delta m_{-}^2=\frac{-i \lambda}{2}\frac{\mu^{4-d}}{\xi^2
 (m_+^2-m_-^2)^2}
\int \frac{d^d p}{(2\pi)^d}\bigg[\frac{i}{p^2-m_-^2+i\epsilon^*(0)}
+\frac{-i}{p^2-m_+^2-i\epsilon^*(0)}\bigg]
\end{eqnarray}

\medskip
\noindent
The  minus in front of the last propagator (of $\varphi_2$)
is explained by  the kinetic term for the ghost field $\varphi_2$ in
 (\ref{action_4g}). The different propagators for $\varphi_{1,2}$
in Minkowski  space-time turn out to become identical in Euclidean
space-time, see last line in (\ref{delta-mass-M}).
Using that under differentiation of (\ref{mpm})
\begin{eqnarray}\label{s3}
\delta m_-^2= \frac{\delta m^2}{\sqrt{1- 4\xi m^2}}\,
\end{eqnarray}
we obtain a result for $\delta m^2$ identical to that in the second
line of eq.(\ref{delta-mass-M}). 
 This confirms that the two descriptions,  in terms
of $\phi$ or of  $\varphi_{1,2}$  are equivalent, as expected.
When $\xi\ra 0$,  $\delta m_-^2\ra\delta m^2$.

We conclude this section by analyzing 
what happens if instead of using the  Lagrangian in 
eq.(\ref{action_4}),(\ref{action_4g}), one insists to 
 use instead its Euclidean version, given below  
\smallskip
\begin{eqnarray}\label{action_E}
\cL_E=\frac{1}{2}\,\phi \,(\,\xi \Box_E^2-\Box_E+m^2\,)\,\phi +\bigg( V_0+
\frac{\lambda'}{4!}\,\phi^4\bigg)
\end{eqnarray}
The one-loop result for $\delta m^2$ computed with the Feynman rules
derived from $\cL_E$ gives:
\smallskip
\begin{eqnarray}\label{E_res}
 -\, \delta m^2& =& 
- \lambda' \,\mu^{4-d} \int_{\bf E} \frac{d^d p}{(2\pi)^d}
\frac{1}{\xi p^4 +p^2+m^2}\nonumber\\[9pt]
&=&\frac{- \lambda'\,\mu^{4-d}}{\sqrt{1-4 \xi m^2}}
 \int_{\bf E} \frac{d^d p}{(2\pi)^{d}}
\bigg[
\frac{1}{p^2+m_-^2}-\frac{1}{p^2+m_+^2}
\bigg]
\end{eqnarray}

\medskip
\noindent
where all $p^2$ are evaluated in Euclidean metric, shown by the
subscript ${\bf E}$. A similar result is found by
working in the corresponding $\varphi_{1,2}$ basis. This result 
has no quadratic divergence but only a logarithmic one.
This would suggest a better UV behaviour of $\delta m^2$ in the Euclidean
 formulation compared to that in (\ref{delta-mass-M}). 
Also, eq.(\ref{E_res}) with $\lambda'\!\ra\! \lambda$
is in contradiction with last line in
 (\ref{delta-mass-M}) which has an opposite sign between the 
 contributions from $m_-$ and $m_+$; this sign was due
to two Wick rotations in opposite directions in 
(\ref{delta-mass-M}),  which brought in an extra (-1) relative sign.
Therefore, the origin of this different result is ultimately due to the analytical
continuation from the Minkowski to the Euclidean space-time,
 not captured by the Euclidean formulation
alone\footnote{To see this in more detail, use in the second line of
eq.(\ref{delta-mass-M}) the relation 
$1/(x-i\epsilon)=1/(x+i\epsilon)+2 i \pi \delta(x)$ with
 $x=p^2-m_+^2$. Then from  eq.(\ref{delta-mass-M}) one obtains a result which
reproduces  the result in the last line of (\ref{E_res}) of the Euclidean theory, 
plus  an extra  correction term, proportional to
$\int d^4 p \,\delta(p_0^2-\vec p^2-m_+^2)$. This  brings a quadratic divergence,
emerging from analytical continuation, not captured by the Euclidean
 theory  alone, of  eqs.(\ref{action_E}), (\ref{E_res}). The
 divergence requires in the end the introduction of a suitable
 operator(s) or set thereof in the  Euclidean theory.}.

The relation at the loop level, between theories formulated in  Euclidean and
Minkowski space-time respectively, is further complicated by the fact
that if the  theory is not renormalisable, a given set of
 operators (counterterms) in Minkowski space-time may correspond
  to a different set of operators in the
  Euclidean spacetime or to the same set  but with
  different coefficients. Also, the Euclidean formulation itself is not 
always  finite. To see these,  add in 
 eq.(\ref{action_E}) the operator $\Delta \cL_E=- z'
 \phi^2\Box_E \phi^2$,
 ($z'>0$) which is dimension 6, like $\phi\Box_E^2\phi$. Its
 effect on  $\delta m^2$ is\footnote{This  operator is
 equivalent to $\Delta \cL_E\!=\!- 2 \phi^3 \Box_E\phi-2 \phi^2
(\partial_\mu \phi)^2$ up to total derivatives and this  is used
 in (\ref{qwqw}). Both operators  in the rhs have similar loop
 corrections to $m^2$, up to overall factors equal to 12 and 4  respectively}
\smallskip
\begin{eqnarray}\label{qwqw}
-\delta m^2\Big\vert_{\Delta \cL_E}=
\frac{-16 z'\,\mu^{4-d}}{\sqrt{1-4 \xi m^2}}\int_{\bf E} \frac{d^d p}{(2\pi)^d}
\bigg[\frac{m_+^2}{p^2+m_+^2} - \frac{m_-^2}{p^2+m_-^2} \bigg]
\end{eqnarray}

\smallskip
\noindent
This result shows that the Euclidean 
theory  has quadratic divergences induced 
by an operator of same dimension as $\phi\Box_E^2\phi$.
Since all operators with same mass dimension should be included
for a comprehensive study, it turns out that even the Euclidean
formulation of a theory with higher derivatives is not necessarily finite.

The sum of the 
radiative corrections of (\ref{E_res}), (\ref{qwqw})
gives a result identical to that in (\ref{delta-mass-M}) of the
Minkowski Lagrangian, provided that $\lambda'\!=\!\lambda/\sqrt{1-4 \xi m^2}$ and
 $z'\!=\!\xi\lambda/(16 \sqrt{1-4 \xi m^2})$. For $\xi\!=\!0$ these relations 
give $\lambda'=\lambda$, $z'=0$ and the two
formulations have identical set of operators and couplings 
 since then the contribution from $m_+^2\sim 1/\xi$ vanishes.
Further, one can also consider
the operator  $\Delta \cL=-z \phi^2\Box \phi^2$  in the Minkowski
Lagrangian eq.(\ref{action_4}) and evaluate its loop corrections to $m^2$;
after doing so, one obtains similar one-loop 
corrections to $m^2$ in the two formulations
from the same set of operators in Minkowski and Euclidean actions
 (up to a redefinition of their coefficients/couplings). This happens
  provided that $z'=(\xi\lambda/16-z)/\sqrt{1-4 \xi m^2}$ and
$\lambda'=(\lambda -64 z m^2)/\sqrt{1-4 \xi\, m^2}$. This may 
 even lead to $z'<0$, for $z>0$, and cause instabilities in the
 Euclidean theory, present in some theories with higher
 derivatives.

To conclude, at the loop level the relation between 
 the Euclidean and Minkowski formulations is more complicated
in the presence of higher dimensional  operators, and in particular
higher derivative ones,  due to  additional
poles induced by  ghost fields, whose  residue  affect the 
analytical continuation. 
In the following, we restrict the analysis to studying the role of
the higher derivative operator only, appearing  in eq.(\ref{action_4}). 
A full analysis including all dimension-six operators and their radiative corrections
is beyond the goal of this paper.

\subsection{The one-loop scalar potential in the presence of higher derivatives}
\label{one-loop-potential}

The  one-loop scalar potential  (vacuum energy) in a 4D  theory  
without higher derivatives  is:
\begin{eqnarray}\label{potential1}
U(\phi)=V(\phi) -\frac{i\,\mu^{4-d}}{2} \int \frac{d^d p}{(2\pi)^d}
\ln \,\frac{p^2 - V'' (\phi)}{p^2 - V''(0)}
\end{eqnarray}
with $V(\phi)$ the tree level potential.
In string theory  the starting point for the vacuum energy
is ultimately a somewhat  similar formula, 
``upgraded''  to respect string symmetries (world-sheet modular 
invariance, etc). In the presence of higher derivative terms,
 eq.(\ref{potential1})  is not valid anymore.

Let us  address the one-loop potential and its renormalisation in the
presence of higher derivatives, using the action in
 eq.(\ref{action_4}). 
In this case the one-loop potential in the Minkowski space is
\smallskip
\begin{eqnarray}\label{potential3}
U(\phi)&=& V(\phi)- \frac{i\,\mu^{4-d}}{2}\int \frac{d^d p}{(2\pi)^d}
\, \ln\frac{p^2- \xi \, p^4 -V''(\phi) + i \epsilon}{
p^2- \xi \, p^4 -V''(0) + i \epsilon}\\[10pt]
&=& V(\phi)-\frac{i\,\mu^{4-d}}{2}
\int \frac{d^d p}{(2\pi)^d}
\,\ln\frac{(p^2 -\alpha_+(\phi)-i\,\epsilon^*(\phi))}{(p^2 -\alpha_+(0)
-i\,\epsilon^*(0))}\frac{(p^2 -\alpha_-(\phi)+i\,\epsilon^*(\phi))}{(p^2
  -\alpha_-(0)+i\,\epsilon^*(0))}
\label{pot3}
\end{eqnarray}
where, as usual $\epsilon>0$, and
\smallskip
\begin{eqnarray}\label{roots}
\alpha_{\pm}(\phi)
\equiv \frac{1}{2 \xi} \Big(1\pm \sqrt{1-4\,\xi\, V''(\phi)}\,\Big),
\quad m^2_\pm=\alpha_{\pm}(0);\quad
\epsilon^*(\phi)\equiv \frac{\epsilon}{(1-4\, \xi\, V^{''}(\phi))^{\frac{1}{2}}}
\end{eqnarray}

\smallskip
\noindent
If $\xi V^{''}(\phi)\ll 1$, 
$\alpha_+(\phi)\approx 1/\xi-V''(\phi)$  (ghost-like part),
and $\alpha_-(\phi)\approx  V''(\phi)$ (particle-like part).

Note the different 
 $\pm i \epsilon^*$ that emerged under the logarithm when
 going from (\ref{potential3}) to (\ref{pot3}), with
 consequences for the analytical continuation to Euclidean space.            
  In many studies, theories with higher derivative operators are
 studied in Euclidean space and  at the end
 it is  {\it assumed} that there exists 
a continuation to the Minkowski space. However, in that case
 ambiguities (regarding which  choice of sign of $i \epsilon$ to
 take) or additional complications  can emerge  when going from 
Euclidean to Minkowski space (see discussion  in the previous section).
 Such issues do not arise when starting with 
the Minkowski formulation eq.(\ref{action_4}) and subsequent
(\ref{action_4g}) and  (\ref{potential3}).
In (\ref{action_4}) the $+i \epsilon$
prescription is consistent with  a well-defined partition function
 for the particle-like degree of freedom
 in Minkowski space, and similar to that  for any scalar  theory 
in the absence
 of higher derivatives ($\xi=0$), considered here a perturbation.
  Since $\phi$  contains a ``piece'' of ghost
(being a ``mixture'' of $\varphi_{1,2}$), it is not surprising that 
this initial prescription  implicitly
fixes the prescription for the ghost-like part as well; this can be
seen  in the complex term under the ``ghost-like''  logarithm in
eq.(\ref{pot3}), which comes with a definite ``prescription'' ($-i \epsilon^*$).

An intriguing aspect that emerged is the $\phi$ dependence
of $\epsilon^*$,  telling us that the condition $\epsilon^*(\phi)$
be very small can be  violated at the dynamical
level. This should
be avoided,  at least because otherwise $\alpha_\pm(\phi)$ 
would have values  closer to each other
and the ghost and the particle reach  masses of similar order of
magnitude.  Then the  theory 
breaks unitarity at a mass scale close to that of the particle and 
this is not something one would  want.
One also needs $\alpha_{\pm}(\phi)$ be real, since otherwise the initial
fourth order theory would have no particle-like degree of
freedom. One would then prefer the ghost have a very large mass,
so that effects associated with  its mass scale are not present
 at low energies. Therefore the particle-like degree
of freedom is  light compared to the ghost-like one, requiring
\smallskip
\begin{eqnarray}\label{cond}
V^{'' }(\phi)\ll \frac{1}{4\,\xi}, \qquad {\rm for \,\,\,any}\quad <\phi>
\end{eqnarray}

\smallskip
\noindent
i.e. the curvature of the potential in ghost mass units be smaller
than unity; this ensures      $\epsilon^*\!\ll\! 1$.

Let us now  discuss the one-loop corrected
 $U(\phi)$ of (\ref{potential3}) which  is  the sum of {\it two}
contributions due to $\alpha_\pm(\phi)$,
each similar to that of a 4D  theory {\it without} the higher
derivative term. Our initial 
 problem of a higher derivative operator in the action 
 is ``unfolded'' into two 4D copies, each with 
its own scalar potential(s) with second derivative(s) $\alpha_\pm(\phi)$.
 After a Wick rotation to the Euclidean
 space, eq.(\ref{potential3}) gives\footnote{To perform the Wick rotations  one
 uses  $\int d^d p \ln\big[p^2-\rho\pm i \,\epsilon\big]
=\pm  i\,\int_{\bf E} d^d p\,\ln\big[p^2+\rho\big]$
To show this, consider  $s>0$ and with $d=4-\epsilon$, then
$\int d^d p \, (p^2-\sigma\pm i\epsilon)^{-s}=\pm i \,(-1)^{-s} \pi^{d/2}
\sigma^{d/2-s}\, {\Gamma[s-d/2]}/{\Gamma[s]}$ ($\sigma>0$).
Here the $+ (-)$  sign is due to an anti-clockwise (clockwise)  Wick
rotation, respectively. One then differentiates this last eq with
respect to $s$ and takes  the limit $s\!\rightarrow\! 0$  to recover the Wick
rotation of the logarithmic term under the integral.}
\medskip
\begin{eqnarray}\label{uofphi}
U(\phi)&=& V(\phi)+\frac{\,\mu^{4-d}}{2}\int_{\bf E} \frac{d^d p}{(2\pi)^d}
\, 
\bigg[\ln\frac{(p^2 +\alpha_-(\phi))}{(p^2 +\alpha_-(0))}
- \ln\frac{(p^2 +\alpha_+(\phi))}{(p^2 +\alpha_+(0))}\bigg]
\end{eqnarray}

\medskip
\noindent
Notice the minus sign above, consequence of the 
analytical continuation, and showing 
the  difference between Minkowski and Euclidean spaces properties
in higher order theories. 
If one started the analysis in an  Euclidean setup instead, by
computing $U(\phi)\sim\Tr(\ln[\xi \Box^2+\Box+V''])$,  one would have 
 instead obtained  a plus sign in front of the last term in
 (\ref{uofphi})!
This difference is due to additional counterterms associated with  
analytical continuation from Minkowski to Euclidean space, as
discussed in the previous section.
As a check of the correctness of our result, if one  takes the first derivative
of $U(\phi)$ with respect to $\phi^2$ at $\phi=0$ one recovers exactly 
$\delta m^2$ of eq.(\ref{delta-mass-M}),
(\ref{delta-mass-E}) in both regularisation
schemes. This is a good consistency check.

Eq.(\ref{uofphi})  tells us something more, when  we consider only 
the {\it field-dependent} part of the integrals, involving $\alpha_\pm(\phi)$.  
Each  of the integrals gives upon integration a
quartic divergence in scale, which may be seen more easily 
in a cutoff regularisation of (\ref{uofphi}). From this equation,
 with a cutoff $\Lambda$
on the above integrals instead of the DR scheme, one has at large
$p^2$
\smallskip
\begin{eqnarray}\label{cnc}
U(\phi)& \sim & \frac{1}{2}\int_{\bf E}^{\Lambda} \!\!
{d^4 p} \,\Big[\ln p^2 + \alpha_{-}(\phi)\,
p^{-2}+ \alpha^2_-(\phi)\,p^{-4}+\cdots\Big]- \Big(\alpha_-\ra \alpha_+\Big)
\end{eqnarray}

\medskip
\noindent
The familiar quartic divergence coming from the integration of $\ln p^2$
and present in the field dependent part of
the particle-like contribution ($\alpha_-(\phi)$)  is cancelled by the similar one
 due to its ghost counterpart, from $\alpha_+(\phi)$. 
This cancellation is due to the minus sign in
(\ref{uofphi}), which is in turn due to the rotation from 
Minkowski to the Euclidean space with opposite prescriptions $\pm i
\,\epsilon^*$ in (\ref{pot3}).  The cancellation is
 similar to that ensured in the presence of softly broken
supersymmetry by\footnote{In
softly broken 
supersymmetric  case one has that $\int\! d^4\! p 
\ln(p^2\!+\!{\rm  M^2})=\sum_J (-1)^{2 J} (2J\!+\!1)\,
{\rm Tr} \!\int\! d^4 p\ln(p^2\!+\!M_J^2)={\rm Str M}^0 
\int\! d^4 p \ln p^2 + {\rm Str M^2}
\!\int\! d^4 p/p^{2}\!-{\rm Str M}^4\int d^4 p/(2 p^{4}) 
+{\rm finite}$ after expanding at large $p^2$;
the first term in the rhs gives ${\rm StrM}^0
\cO(\Lambda^4\ln\Lambda)$; in our case this UV cutoff dependent
term is absent in the   field dependent part alone,
due to a cancellation between
particle and its ghost contribution. }
 equal bosonic and fermionic degrees of freedom ${\rm Str} M^0=0$.
Note however the presence of a   UV-finite, cutoff-independent correction,
quartic in  the mass of the 
ghost (i.e. in the scale of the higher derivative operators);
this comes from $\alpha_+^2(\phi)\sim 1/\xi^2\equiv M_*^4$ in the limit of small
$\xi$. This discussion  shows that ultimately  higher derivative terms 
 play a role in addressing the cosmological constant problem.
These observations and the results of eqs.(\ref{action_4g}),
(\ref{pot3}), (\ref{uofphi}), (\ref{cnc})
have some
similarities to the  proposal in \cite{Kaplan:2005rr} for the
cosmological constant.

We now proceed to address the renormalisation of $U(\phi)$.
 Eq.(\ref{uofphi}) gives in the DR scheme:
\medskip
\begin{eqnarray}
U(\phi)&=& V(\phi)+ \Big[\Delta U_1(\phi)+\Delta
U_2(\phi)\Big]- \Big[\Delta U_1(0)+\Delta U_2(0)\Big]
,\label{eq42}
\\[10pt]
\Delta U_1(\phi) &=&
c_o \,\bigg[\frac{1}{\omega}+c_1 \bigg]
\bigg[\alpha^2_-(\phi)-\alpha^2_+(\phi)\bigg],\qquad \omega\equiv
4-d\ra 0;
\label{eq43}
\\[10pt]
\Delta U_2(\phi)&=&
\frac{c_o}{2}\,\bigg[
-\alpha_-^2(\phi)\ln \frac{\alpha_-(\phi)}{\mu^2}
+\alpha_+^2(\phi)\ln \frac{\alpha_+(\phi)}{\mu^2}
\bigg], 
\end{eqnarray}

\medskip
\noindent
where
$c_o\equiv - 1/{2 \,(4\pi)^2}$,\,\, $c_1\equiv {1}/{4}
\ln\big[(4\pi)^2 e^{3-2\gamma}\big]$.

The contribution $\Delta U_1$ involves a square root of the fields,
which  will require us to introduce infinitely many 
counterterms, already at the one-loop level. This may be seen from
the following equation, valid for 
 $\xi V''(\phi)\!\ll\! 1$:
\medskip
\begin{eqnarray}
&&\alpha^2_-(\phi)- \alpha^2_+(\phi)
= \,\frac{-1}{\xi^2}\,\sqrt{1- 4\,\xi\,V^{''}(\phi)}\nonumber\\[9pt]
&&\quad\quad =\,\,
-\frac{(1-4 \xi m^2)^{\frac{1}{2}}}{\xi^2}
+\frac{\lambda \phi^2}{\xi\,(1-4 \xi  m^2)^{\frac{1}{2}}}
+\frac{\lambda^2\,\phi^4}{2 (1-4 \xi m^2)^{\frac{3}{2}}}
+\frac{\lambda^3\,\xi\,\phi^6}{2 (1-4 \xi m^2)^{\frac{5}{2}}}
+\cO(\xi^2 \phi^8)\quad
\label{last}
\end{eqnarray}

\medskip
\noindent
where we used $V(\phi)$ of (\ref{action_4}).
 The series above 
requires one introduce {\it higher dimensional} counterterms.
However, after renormalisation,  under the condition
 $\xi V''(\phi)\!\ll\! 1$, the usual 4D counterterms are sufficient and
the theory ``appears'' as 4D renormalisable. This condition is respected 
when the scale of higher derivative
operators, $M^2_*\equiv 1/\xi$ is high enough and when the vev of the
field $<\phi>$ has no runaway to infinity. This is exactly our initial
condition (\ref{cond}) imposed on physical arguments such as absence
of unitarity violation at low scales,  etc.
In the  approximation of neglecting the $\phi^6$ counterterm and
higher ones, the renormalised potential $U_r$ is
\smallskip
\begin{eqnarray}\label{c-terms}
U_r(\phi)& =& U(\phi)+\delta V_0+\frac{\delta m^2}{2}\,\phi^2 +
\frac{\delta \lambda}{4!}\,\phi^4\\[8pt]
\frac{\delta\,m^2}{2}  &=&   -  c_o\, \lambda\,
\,\bigg[\frac{1}{\omega}+c_1 \bigg] \,
\frac{1}{\xi\,(1-4\, \xi \,m^2)^{\frac{1}{2}}}
+\frac{a_0}{2} \,m^2, \\[8pt]
\frac{\delta\lambda}{4!}& = & -
c_o\,\lambda^2\bigg[\frac{1}{\omega}+c_1 \bigg]
\,\frac{1}{2\,(1-4\,\xi\,m^2)^{\frac{3}{2}}} 
+\frac{\lambda}{4!}\, b_0,\qquad
\delta V_0  = -V_0\label{renorm}
\end{eqnarray}

\medskip
\noindent
The cosmological constant term $V_0$ only undergoes a 
finite renormalisation. The mass undergoes renormalisation
from both the particle  and the higher derivative 
term (the ghost),  which are 
contributing to the quadratic
divergence of the mass. This is seen 
from the $\xi$ dependence  of  $\delta m^2$; in 
 ordinary $\phi^4$ theory, the $\xi$-dependent factor is replaced by $m^2$.
  The coupling constant is also renormalised,
despite the presence of the higher derivative term in the action.
The coefficients   $a_0$ and $b_0$ account for any finite
 part of the counterterms, i.e. are  regularisation scheme dependent
constants,  which can be fixed  by suitable 
 normalisation conditions.\footnote{These are 
$m^2=\frac{\partial^2 U_r(\phi)}{\partial\phi^2}\big\vert_{\phi=0},$\,\,
$\lambda=\frac{\partial^4 U_r(\phi)}{\partial\phi^4}\big\vert_{\phi=0}$}
The result is
\medskip
\begin{eqnarray}\label{final-one-loop}
U_r(\phi)=\bigg[\lambda-
\frac{\partial^4\Delta
  U_2}{\partial\phi^4}\bigg\vert_{\phi=0}\bigg]
\frac{\phi^4}{4!}
+\frac{1}{2} \bigg[m^2-\frac{\partial^2\Delta
U_2}{\partial\phi^2}\bigg\vert_{\phi=0}\bigg]\phi^2+\Delta
U_2(\phi)-\Delta U_2(0)
\end{eqnarray}

\medskip
\noindent
This is the one-loop  scalar  potential in the theory 
with higher derivative term (the  derivatives are given in the
Appendix eq.(\ref{eqI})).
In this very  minimal case there is no spontaneous symmetry breaking,
(first derivative vanishes  only for $\phi=0$), 
but as mentioned, in  models which include 
additional interactions, fermions, etc., this may be possible.
The above form of the one-loop potential can be used in such models. 
Note that the $\alpha_+$ dependent part (ghost contribution) 
in the last two terms of $U_r$
is, at small~$\xi$ 
\begin{eqnarray}
\Delta U_2(\phi)-\Delta U_2(0)=
\alpha^2_+(\phi)
\ln\frac{\alpha_+(\phi)}{\mu^2}-\alpha^2_+(0)
\ln\frac{\alpha_+(0)}{\mu^2}\!+\!\cO(\xi^0)
=
\frac{\lambda \phi^2}{2\,\xi} (2 \ln\xi-1)\!+\!\cO(\xi^0)
\end{eqnarray}

\medskip
\noindent
where we ignored the $\alpha_-$ (particle-like) part which cannot
introduce singular terms if $\xi\ll 1$. Note that the quartic
mass dependence $1/\xi^2=M^4_*$ present in the field dependent part
(from $\alpha^2_+(\phi)$) is
cancelled by that from $\alpha_+(\phi=0)$. This leaves only a term 
proportional to $\phi^2/\xi$, which is only  quadratic in the (high) mass of the
ghost\footnote{
It is for this cancellation to take place that we kept the constant
$\Delta U_2(0)$ in the potential, rather than introduce it 
in the renormalisation of $V_0$ in (\ref{c-terms}). 
This then avoids the fine tuning of the 
tree level $V_0$ to cancel  one-loop  $\Delta U_2(0)\sim 1/\xi^2\sim M^4_*$ in the
limit of decoupling the ghost i.e. at large $M_*$, see also eq.(\ref{renorm})}.
At non-zero $\phi$ the ghost contribution does not decouple
in the potential when $\xi\ra 0$. For smaller $\xi$ 
the potential  acquires a steeper dependence on $\phi$;
at large $\xi$ the higher dimensional terms neglected so far become 
 important.

\bigskip
\subsection{O'Raifeartaigh supersymmetry breaking with higher derivatives.}\label{O'R}

We shall now address the implications of the higher derivative operators
in a 4D N=1 supersymmetric context. We  consider first the case of 
 O'Raifeartaigh model with additional, supersymmetric higher derivative terms
and  spontaneous supersymmetry breaking. We  compute
 the one-loop correction to the self-energy of a scalar field in the
presence of higher derivative operators and investigate its UV behaviour.
 The action   is:
\medskip
\begin{eqnarray}\label{O'Rmodel}
\cL
&=&\!\int d^4\theta \,\sum_{j=0}^2 \Phi_j^\dagger \Big(1+ \xi_j \,\Box\Big)\Phi_j
+\bigg\{\int d^2\theta \bigg(\lambda\,\Phi_0 \, \Phi_2 + 
g\,(\Phi_0^2-M^2)\, \Phi_1 \bigg)+ c.c.\bigg\}\nonumber\\[7pt]
&=&
\sum_{j=0}^2 F^*_j \Big( 1+\xi_j {\Box} \Big) \,F_j
- \varphi^*_j\Box \Big( 1+\xi_j {\Box} \Big) \,\varphi_j
+i\, \partial_\mu \bar \psi_j  \,\bar\sigma^\mu
\Big( 1+\xi_j {\Box} \Big)  \psi_j-\Big(g\,M^2 F_1+c.c.\Big)\nonumber\\[9pt]
&+& \Big[
\lambda \Big(\varphi_0 F_2 +F_0\varphi_2-\psi_0\psi_2\Big)
+g\Big(2\varphi_1 \,\varphi_0\, F_0+F_1\,\varphi_0^2 -2\, \varphi_0\,
 \psi_1\,\psi_0 -\varphi_1
\,\psi_0\,\psi_0\Big)+c.c.\Big]\quad
\end{eqnarray}

\medskip
\noindent
where the chiral superfield $\Phi_j$ has components
$\Phi_j=(\varphi_j, \psi_j, F_j)$.
We use standard conventions\footnote{We
  use the notation $\sigma^\mu=(\sigma^0,\sigma^i)$
where $\sigma^i$ are Pauli matrices, with 
$\sigma^0=1_{2\times 2}$; its elements are labelled  $\sigma^\mu_{A\dot A}$;
also $\overline\sigma^\mu\equiv (\overline\sigma^0,\overline
\sigma^i)=(\sigma^0,-\sigma^i)$ whose  elements are
$(\overline\sigma^\mu)^{\dot A A}=\varepsilon^{AB} \varepsilon^{\dot A \dot
  B} \sigma_{B \dot B}^\mu, \,\,\,A=1,2; \dot A=\dot 1,\dot 2$;
also $\varepsilon_{11}=\varepsilon_{22}=0$,\,
$\varepsilon_{12}=-1=-\varepsilon_{21}$,
$\epsilon^{AB}=\epsilon_{AB}^T$ with similar definitions for ``dotted''
$\varepsilon$; finally
$tr (\sigma^\mu\overline \sigma^\nu)=2\eta^{\mu\nu}, \eta^{\mu\nu}=(+,-,-,-)$.
} and for any Weyl spinors $\psi\psi\equiv
\psi^A\psi_A$, $\overline\psi\overline\psi\equiv
\overline\psi_{\dot A}\overline\psi^{\dot A}$ and also
$\overline\psi \,\overline\sigma^\mu\,\psi\equiv
\overline\psi_{\dot A}\, (\overline\sigma^\mu)^{\dot A B} \,\psi_B$.
The terms involving the $\Box$ operator are  manifest supersymmetric
if we recall that $-1/16 \,{\overline D}^2 \, D^2 \,\Phi_j=\Box \,\Phi_j$, 
for $\Phi_j$ a left chiral superfield, and where $D_A, \,\overline
D_{\dot A}$ are
supersymmetric covariant derivatives.

If all $\xi_j=0$, $j=0,1,2$ one recovers the
familiar O'Raifeartaigh model of supersymmetry breaking. 
We review this briefly, before returning to the case of non-zero
$\xi_j$.
In this model, take  $g, M,\lambda$   all  real  and non-zero; then
the  model has spontaneous supersymmetry breaking,  since the potential $\cV$ cannot
vanish under this assumption. Indeed, the potential is
\medskip
\begin{eqnarray}\label{vv}
\cV=\sum_{i=0,1,2} \vert F_i\vert^2=
\vert\,\lambda \,\varphi_2 +2 \,g\,\varphi_1\,\,\varphi_0\,\vert^2
+
\vert \,g\,(\varphi_0^2-M^2)\,\vert^2
+
\lambda^2 \,\vert \,\varphi_0\,\vert^2
\end{eqnarray}

\medskip
\noindent
where $F_0, F_1, F_2$ are given by the three terms above, in this
order. The condition $\cV=0$ has no solution if  all $g,
M, \lambda$ are non-zero; the minimum conditions $\partial
\cV/\partial \varphi_{1,2}=0$ when satisfied, give 
an extremum value:
$\cV_{m}=\lambda^2 \,\vert \,\varphi_0\,\vert^2 +\vert
\,g\,(\varphi_0^2-M^2)\,\vert^2$; further, from  $\partial \cV_m/\partial
\varphi_0=0$ one obtains that $\varphi_0$ is real and that  $\varphi_0 \,(\lambda^2+2 g^2
(\varphi_0^{2}-M^2))=0$. For this equation there are two possibilities: {\bf (a):}~if 
 $\lambda^2\geq  2 g^2 M^2$  then $\varphi_0=0$ and then
$\cV_{m}(\varphi_0\!=\!0)=g^2 M^4$ is the minimum of the potential 
and  spontaneous supersymmetry
breaking takes place. In this case $F_0=F_2=0$, $F_1=g\,M^2$.
{\bf (b):}~if  $\lambda^2< 2 g^2 M^2$, one has  
two  non-zero roots from:  $2 g^2\varphi_0^2=
2 g^2 M^2-\lambda^2$ which correspond to the minimum of the
potential, while $\cV_{m}(\varphi_0=0)$ is now a local maximum. In 
this case $\cV_{m}=g^2 M^4+
\varphi_0^2 (\lambda^2 -2 g^2 M^2)+g^2 \varphi_0^4$ and therefore, in addition to
 spontaneous supersymmetry breaking, there
 is internal (spontaneous) symmetry breaking  with respect to $\varphi_0$
 (negative ``mass'' $\lambda^2 -2 g^2 M^2$). 
The  symmetry of the potential which is
broken is a $Z_2$ symmetry, $\varphi_0\ra -\varphi_0$. 
In this case, $F_0=0$, $F_1=g(M^2-\varphi_0^2)$, $F_2=-\lambda\varphi_0$
where $\varphi_0$ denotes the above non-zero roots.
In both cases discussed above, one notices that 
the condition $g M^2=0$ restores 
supersymmetry and a vanishing $\cV$.
 This concludes the review of the O'Raifeartaigh model.

We return now to the action in (\ref{O'Rmodel})
and consider the presence of higher derivatives for some or  all
superfields. If $\xi_j\not\! =\!0$ for some $j$
 the corresponding   $F_j$ field is dynamical and has  a
ghost field partner ($\Box F_j$). The off-shell counting of the
bosonic and fermionic real degrees of freedom works as in the absence 
of the higher derivatives,  but now each field
has a  counterpart  (ghost): $\varphi_j$ (2), $\Box
\varphi_j$ (2), $F_j$ (2), $\Box F_j$ (2), $\psi_j$ (4), $\Box \psi_j$ (4), 
since $\Box f$, where $f$ is some field, is seen
as an extra degree of freedom, although not
 independent of $f$  (see  discussion around eq.(\ref{action_4g})).

In the presence of the higher derivatives ($\xi_j\not=0$), the
effective  potential at the tree level has a minimum which is not
affected by the fact that  $F_j$ has now a dynamical nature.  Indeed, the
extremum  condition $\partial\cV/\partial F_j=0$,  where $\cV$ is
obtained from  (\ref{O'Rmodel}), provides solutions
for $F_j$, regardless of whether $F_j$ is dynamical or not.
After inserting back these  values of $F_j$  in the
potential, one obtains for $\cV$ at {\it the extremum point,} 
a result identical to (\ref{vv}), and with squared
absolute values  of $F_{0,1,2}$ respectively given by the three terms
in (\ref{vv}),  in this order. With this observation, the discussion of
supersymmetry breaking (at the tree level) in the presence of higher derivatives
follows identically that after eq.(\ref{vv}) in the absence of higher
derivative terms. As before,  if $g M^2=0$ then supersymmetry is restored.

In this framework we  now investigate the UV behaviour of the self
energy of scalar field $\varphi_0$  in the presence of the higher
derivative terms. This is  interesting since the supersymmetry  breaking is 
spontaneous, and we would like to
see the dependence of the quantum corrections to the mass of
$\varphi_0$, on the scale of (supersymmetric) higher
derivative terms ($1/\xi_i$) and on the UV cutoff ($\Lambda$). 
Does spontaneous supersymmetry breaking remain soft in the presence of
higher derivative terms, and if so,  
under what conditions? These questions are addressed below.

\begin{figure}[t]{
\centerline{\psfig{figure=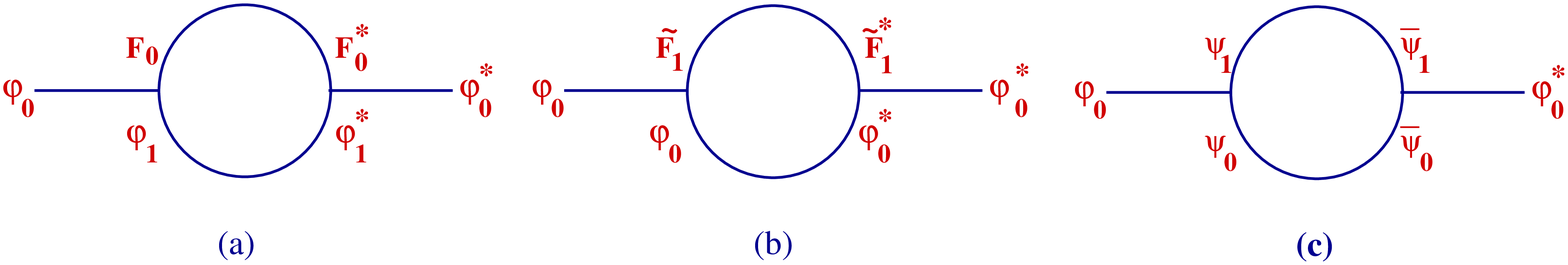,height=1.in,width=6.2in,angle=0}}
\def\baselinestretch{1.1}
\caption{\small The  Feynman diagrams (off-shell) which contribute 
to the self-energy of  $\varphi_0$ in Section \ref{O'R}.} 
\label{fig2}}
\end{figure}

To compute the loop corrections to 
the mass of $\varphi_0$ (see Fig.~\ref{fig2})
 we first  need to replace in  (\ref{O'Rmodel}),
  $F_1\ra \tilde F_1+g M^2$.
After this, the new Lagrangian equals  $\cL'(F_1\ra \tilde F_1)+\Delta \cL$ 
where $\cL'$ is that of  eq.(\ref{O'Rmodel}) 
but without  the linear term in $F_1$, and
\smallskip
\begin{eqnarray}\label{newL}
\Delta\cL= g^2 M^2 \varphi_0^{* 2}+ g^2 M^2 \varphi_0^{2}
-g^2 M^4   
\end{eqnarray}
In the presence of higher derivative terms one 
must pay attention to the prescriptions for the propagators' poles.
It is easier to understand these by first considering a simpler case,
of $\lambda=0, M=0$. In this case
\smallskip
\begin{eqnarray}\label{aa1}
<\tilde F_1\,\tilde F_1^*>= \frac{i
}{\xi_1\,(\Box+ 1/\xi_1  -i\,\tilde \epsilon)},\qquad
<F_j\, F_j^*>=\frac{i
}{\xi_j\,(\Box+ 1/\xi_j-i\,\tilde \epsilon)},\qquad j\equiv 0,2
\end{eqnarray}
Also
\begin{eqnarray}\label{aa2}
\qquad 
<\varphi_j\,\varphi_j^*>&=&\frac{-i}{\Box (1+ \xi_j \Box)
  -i\epsilon}=\frac{-i}{(\Box-i\epsilon)(1+\xi_j\,\Box+i\,\epsilon\,\xi_j)}
,\quad\quad j\equiv 0,1,2.
\end{eqnarray}

\medskip
\noindent
which can further be written as a difference of a particle-like and
ghost-like propagators. Finally
\smallskip
\begin{eqnarray}\label{fermionpropagators}
<\psi^A_j \overline \psi^{\dot B}_j > & = & 
 \,\sigma^{\mu\,\,A \dot B} \,\, \partial_\mu\,
\frac{1}{\Box (1+\xi_j \,\Box)-i\tilde{\tilde\epsilon}}, \qquad j=0,1,2.\quad
\end{eqnarray}

\smallskip
\noindent
while  $<\overline \psi_{\dot A} \psi_{B} >$ has a form similar 
to that of $<\psi^A \overline \psi^{\dot B} >$,
 but with $\sigma^{\mu\,\,A \dot B}$ replaced by
$\overline\sigma_{\dot A  B}^{\nu}$.
Here we used the indices $A, B=1,2$, $\dot A, \dot B=\dot 1, \dot 2$. 
To obtain the momentum 
representations one  replaces 
$\Box\equiv \partial_\mu \partial^\mu
\ra -p^2$ and $\partial^\mu \ra - i p^\mu$.

The $ i \,\epsilon$ prescription for the scalar fields is rather
standard, as in the absence of higher derivative operators, and this
was discussed in the previous section; 
the same must be true for fermions, and the prescription we took for
their propagator  is consistent with a 
$+i\tilde{\tilde\epsilon}\,\psi\,\psi$ shift
 present in the Minkowski partition function exponent, to ensure its
convergence; therefore we have  $\tilde{\tilde \epsilon}=\epsilon$ ($\epsilon>0$);
 this is also  confirmed later by supersymmetry arguments; 
however, to be more general, let us keep the sign of
 $\tilde{\tilde \epsilon}$ arbitrary.
For the auxiliary  fields propagators,  comparison 
of the higher derivative terms for $F_j$ in (\ref{O'Rmodel}) against
the second term in (\ref{action_4g}) suggests that  $\tilde
\epsilon =-\epsilon$ ($\epsilon>0$), but to keep track of its
effects,  let us keep  the sign of $\tilde \epsilon$  arbitrary, too.

Returning to the general case of non-zero $\lambda$ and $M$, the 
 propagators  change now, due to
 mass-mixing terms, proportional to $\lambda$, $M$, see (\ref{O'Rmodel}), (\ref{newL}).
An exception is $<\tilde F_1\,\tilde F_1^*>$ which remains unchanged.  
For the new propagators that we need in Figure~\ref{fig2}, the above prescriptions
still apply and are used for the ``diagonal entries'' in the
relevant matrices, as described below. More explicitly,
  $<F_0 F_0^*>$ is now given by the element $(\cN^{-1})_{44}$ 
where $\cN^{-1}$ is the inverse of matrix $\cN$ defined by
 $\cL\supset (1/2)\, \,\vec w \,\cN\,{\vec w}^{*\, T}$ with $w$  the vector 
basis $w\equiv (\varphi_2^*, \varphi_2, F_0^*, F_0)$.
Similarly, the new propagator  $<\varphi_0\,\varphi_0^*>$ 
is given by the entry $(\cM^{-1})_{22}$, where the matrix $\cM$ is read
 from   $\cL\supset (1/2)\,\, \vec\gamma\,
 \cM\,{\vec \gamma}^{*\, T}$, with $\vec\gamma\equiv (\varphi_0^*,
 \varphi_0, F_2^*, F_2)$.  Finally, 
 $<\overline\psi_0\,\psi_0>$ is now given by the element
 $(\cP^{-1})_{11}$ of the inverse matrix $\cP^{-1}$ where $\cP$ can be read
from $\cL\supset  (1/2)\,\,\vec\delta \,\cP\, {\vec \delta}^{*\, T}$
 with $\vec \delta \equiv (\overline \psi_0,\psi_0, \overline \psi_2,
 \psi_2)$.
The detailed  expressions of these propagators found as described
 here, are  given in 
the Appendix, eqs.(\ref{prop1}) to (\ref{prop3}); these
 expressions recover, 
if  $\lambda\ra 0, M\ra 0$, those quoted in eqs.(\ref{aa1}) to
 (\ref{fermionpropagators}), including the prescriptions for the poles.

Using eqs.(\ref{prop1}) to (\ref{prop3}), one can evaluate the
diagrams which contribute to the mass of $\varphi_0$ (Figure~\ref{fig2}). 
The results are\footnote{Any $\epsilon$-prescriptions in the
  numerators of integrals/propagators are irrelevant and should be set to zero whenever
  present below; they are shown  only to help trace
  the origin of corresponding terms in component formalism.}

\begin{eqnarray}\label{wrwr}
(a)\!\!\!&=&\!\!\!
(2 i g)^2\! \int\! \frac{d^4 p}{(2\pi)^4}
\frac{(-1)\,\,(p^2\,(1-\xi_2 p^2)+i\,\epsilon\,)}{
(p^2\,(1-\xi_1\,p^2)+i\epsilon)\,\Big[
\,(1-\xi_0\,p^2-i\,\tilde 
 \epsilon\,\xi_0)\,\,(p^2 (1-\xi_2\,p^2)+i \epsilon) -\lambda^2
\Big]}\nonumber
\\[11pt]
(b)\!\!\!&=&\!\!\!\! \frac{4 \,(i\,g)^2}{2}\!\!\int\!\! \frac{d^4
  p}{(2\pi)^4}
\frac{-(1-\xi_2\,p^2-i\,\tilde\epsilon\,\xi_2)}{
(1-\xi_1\,p^2-  i\,\tilde \epsilon\,\xi_1)
\,\Big[(1-\xi_2\, p^2-i\,\tilde\epsilon \,\xi_2)\,
\,(\,p^2\,(1-\xi_0\, p^2) +\rho +i\,\epsilon)\!-\!\lambda^2\Big]}_{\!\rho=2 g^2
  M^2}
\nonumber\\[6pt]
&&\qquad \qquad\quad 
+\,(\rho\ra -2 g^2 M^2)
\nonumber\\[11pt]
(c)\!\!\!&=&\!\!\! (2 i g)^2\! \int \!\frac{d^4 p}{(2\pi)^4}
\frac{2 \,p^2  \,(p^2(1-\xi_2\,p^2)+i\tilde{\tilde\epsilon})}
{(\,p^2\,(1-\xi_1\,p^2) +i\,\tilde{\tilde \epsilon})\,\,
\Big[ \,( p^2(1-\xi_2\,p^2)\!+\!i\tilde{\tilde\epsilon})\,\,
( p^2(1-\xi_0\,p^2)\!+\!i\tilde{\tilde\epsilon})-\lambda^2\,p^2\,]\,\Big]}\quad\,\,
\end{eqnarray}

\medskip
\noindent
 We used $tr\,\,[\sigma_\mu\,\bar\sigma_\nu]
\partial_\mu\,\partial_\nu=2\,\Box$
and that diagram (b) has a symmetry factor of~4.

For simplicity   consider in the following the case when
$\xi_0=\xi_2=0$, so only the $\Phi_1$ superfield has a higher derivative
term in (\ref{O'Rmodel}). 
We also assume that $1/\xi_1\not = \lambda^2$, $1/\xi_1\not =
\lambda^2-2 g^2 M^2$. (In fact
one should take $\xi_1\lambda^2\ll 1$, $\xi_1 g^2 M^2\ll 1$ on
 grounds similar to those discussed in the non-supersymmetric case;
 here $\lambda$ and $g^2\,M^2$ play a role similar to $m$ in
 Section~\ref{non-susy}). One obtains\footnote{In expressions (a), (c) we use
 $p^2(1-\xi_j   \,p^2)+i\,\epsilon=-\xi_j\,(p^2+i\,\epsilon)\,(p^2-1/\xi_j-i\,\epsilon)+
\cO(\epsilon^2)$,  (see also (\ref{aa2}) showing  its origin). This also allows
for a check of the  cancellation, for exact supersymmetry,
 of the corrections in (\ref{wrwr})}
\smallskip
\begin{eqnarray}\label{www}
(a)&=&
\,\frac{(2\,i\,g)^2}{1-\xi_1\,\lambda^2}\int  \frac{d^4 p}{(2\pi)^4}
\,\,\bigg[\frac{1}{p^2-1/\xi_1-i\,\epsilon}-\frac{1}{p^2-\lambda^2+i\,\epsilon}\bigg]
\nonumber\\[12pt]
(b)&=&
\frac{(2\,i\,g)^2}{1-\xi_1\,\lambda^{'\, 2}}\int  \frac{d^4 p}{(2\pi)^4}
 \bigg[\frac{1}{p^2-1/\xi_1+i\,\tilde\epsilon}
-\frac{1}{p^2-\lambda^{'\,2}+i\,\epsilon}\bigg]\,\frac{1}{2}
+\Big(\lambda^{' 2}\ra \lambda^{'' 2}\Big),
\nonumber\\[12pt]
(c)&=&
\frac{(2\,i\,g)^2}{1-\xi_1\,\lambda^{2}}
\int\frac{d^4 p}{(2\pi)^4}
\bigg[\frac{-2}{p^2-1/\xi_1-i\,\tilde{\tilde\epsilon}}-\frac{-2}{p^2-\lambda^{2}+2
    i\,\tilde{\tilde\epsilon}}\bigg]
\end{eqnarray}

\medskip
\noindent
where we introduced $\lambda^{'\,2}\equiv\lambda^2-2g^2 M^2$ and
 $\lambda^{''\,2}\equiv\lambda^2+2g^2 M^2.$ 

If $\xi_1=0$ then the usual cancellations for spontaneously broken  supersymmetry are
present, with no quadratic divergences, and only logarithmic terms (in UV cutoff) present. 
To see this, note that in this limit the first term in every square
bracket in (\ref{www}) vanishes, and any contributions from corresponding
 ghost fields are decoupled.  The second term in any of the above brackets
represents the usual contribution in the absence of higher derivative
terms, with overall $\xi_1$-dependent 
coefficients equal to unity when $\xi_1=0$. This supersymmetric
cancellation is  
consistent with the prescription $\tilde{\tilde\epsilon}>0$ and
$\tilde \epsilon<0$ as stated after eq.(\ref{fermionpropagators}).

For non-zero $\xi$, the ghost-like fields contribute to the radiative
corrections and to their UV behaviour; also, in this case the coefficients of the 
``normal'' contributions (for $\xi_1=0$)  are now multiplied by
$\xi$-dependent factors, different from 1. In this case,
and using  the notation $\tilde \epsilon=v\,\epsilon$, $\tilde{\tilde
  \epsilon}=u\,\epsilon$, (where $v=-1,\, u=1$), the result of adding the (a),
(b), (c) contributions  is\footnote{One has
$$\int d^4p \,\,\frac{1}{p^2-m^2\pm i\,\epsilon}=\mp\, 
i\,\pi^2\,\Big[\,\Lambda^2-m^2\,\ln [1+\Lambda^2/m^2]\,\Big],\qquad
\epsilon>0$$ 
If $m^2\gg \Lambda^2$, as it may happen when we decouple
the ghost, $m^2\sim 1/\xi$, then the quadratic divergence disappears.}

\medskip
\begin{eqnarray}\label{eqeq}
\Delta m^2_{\varphi_0}
&=&\frac{g^2}{4\pi^2}\,\bigg\{ 
\bigg[ \frac{2(1-2 u)}{1-\xi_1\,\lambda^2} + \frac{1-v}{2( 1-\xi_1\,
     \lambda^{' \,2})}
 + \frac{1-v}{2( 1-\xi_1\,
     \lambda^{'' \,2})}
\bigg]\,\Lambda^2+
 \frac{ (1-2 u)}{1-\xi_1\lambda^{2}}\,\,g(\Lambda^2,\lambda^{2})
\nonumber\\[12pt]
&+&
\bigg[\frac{1-2\,u}{1-\xi_1\lambda^2}-
\frac{v/2}{1-\xi_1\,\lambda^{'\,2}}-\frac{v/2}{1-\xi_1\,\lambda^{''\,2}}
\bigg] \,
g(\Lambda^2,1/\xi_1)
+\bigg[
\frac{g(\Lambda^2,\lambda^{'\,2})}{2\,( 1-\xi_1\lambda^{'\,2})}
+(\lambda'\ra \lambda^{''})\bigg]
\quad
 \end{eqnarray}

\medskip
\noindent
 with $g(\Lambda^2,m^2)=-m^2 \ln(1+\Lambda^2/m^2)$.
The above expression becomes (with $v=-1,\, u=1$) 

\smallskip
\begin{eqnarray}\label{quadratic}
\Delta m^2_{\varphi_0} &=&
\frac{g^2}{4\pi^2}
\frac{2\, \xi_1^2 \,(2 g^2 M^2)^2}{(1-\xi_1\,\lambda^2)\,
(1-\xi_1\,\lambda^{'2})\, (1-\xi_1\,\lambda^{'' 2})}
\,\,\Lambda^2+\cO(\ln \Lambda)
\end{eqnarray}

\medskip
\noindent
with the notation $\lambda^{'\,2}\equiv\lambda^2-2g^2 M^2$ and
 $\lambda^{''\,2}\equiv\lambda^2+2g^2 M^2.$ 

Before addressing the result obtained, let us remind  that
the ``prescriptions'' we took $v=-1$, $u=1$ or equivalently
$\tilde\epsilon=-\epsilon$ and $\tilde{\tilde\epsilon}=\epsilon$,
are those stated in the text after eq.(\ref{fermionpropagators}),
 and it is re-assuring  to know that they are also consistent with
 supersymmetry arguments and cancellations that take place for exact 
supersymmetry, as discussed earlier.
One may ask whether the other  ``possibilities''
 could  be correct, such as $v=1, \,u=\pm 1$ or $v=-1, u=-1$? The
 answer is negative, and  they can be
easily ruled out; for example,  if supersymmetry is exact ($M=0$)
or only softly broken but without higher derivatives ($\xi_1=0$),
 the sum of quantum corrections would still have quadratic divergent
 terms for these 
``choices'', which is clearly not allowed\footnote{If $ v=1$, $u=\pm 1$, the 
quadratic term is $\frac{g^2}{4\pi^2}\frac{ 2 (1-2 u)
\Lambda^2}{1-\xi_1 \lambda^2}$; if $v=u=-1$ the quadratic term is
proportional to 
$\Lambda^2 (4 (1-2 \xi_1\lambda^2)-3 \rho^2 \xi_1^2 +4
 \lambda^4 \xi_1^2)$,  with $\rho\!=\!2 g^2 M^2$;
 these do not  vanish when restoring supersymmetry (by taking M=0) or if
 $\xi_1\!\ra\! 0$.}.

According to (\ref{quadratic}), we find the interesting result
that  supersymmetry breaking, although spontaneous, can nevertheless
bring in  quadratic divergences at one-loop, if higher
derivative supersymmetric terms are present in the initial action. 
This result is not in contradiction with soft supersymmetry breaking
theorems \cite{GG}, which do not include higher dimensional
derivative supersymmetric terms.

The quadratic divergence found for $\Delta m_{\varphi_0}^2$ 
  has a coefficient that is  proportional to 
$g^4 M^4$ which is related
to the ``amount'' of supersymmetry breaking, and inverse proportional
to the mass scale of the higher derivative operators $M_{1,*}^2\equiv 1/\xi_1$.
If $\xi_1$ is small enough i.e. the scale of higher derivative operator
is high (required in the end  for model building, phenomenology, reasons of 
unitary, etc), the quadratic divergence has a smaller coefficient, 
 but it is still present.
 In the special limit of decoupling  the higher derivative 
operators ($\xi \,\Lambda^2 \ll 1$) we recover the usual result that no quadratic 
divergences are present in spontaneous supersymmetry breaking, in the
  absence of higher derivative terms in the action.

Another case one can  consider is  $\xi_0=\xi_1=\xi_2\equiv \xi$, when all fields
in the model have higher derivative operators, and all auxiliary fields
are dynamical. The one-loop correction of eq.(\ref{wrwr}) becomes
\medskip
\begin{eqnarray}\label{OR2}
(a)&=&\!(2 i g)^2 \int \frac{d^4 p}{(2\pi)^4}
\frac{-1}{\Big(1-\xi\,p^2-i \,\xi\, \tilde\epsilon\,\Big)\,
\Big(\,p^2\,(1-\xi\,p^2)+i\,\epsilon\Big)-\lambda^2}\nonumber
\\[11pt]
(b)\!\!&=&\!\!\frac{(2 i\,g)^2}{2}\!\!\int \frac{d^4 p}{(2\pi)^4}\!
\frac{-1}{\Big(1-\xi p^2-i \,\xi\, \tilde\epsilon\,\Big)\,
\Big[ (p^2\,(1-\xi\, p^2) +i\epsilon)+\rho\,\Big]-\lambda^2}
\bigg\vert_{\rho=2 g^2 M^2}\!\!\!\!
+\Big[\rho\ra -2 g^2 M^2\Big]\nonumber\\[11pt]
(c)&=&(2 i\,g)^2\int \frac{d^4 p}{(2\pi)^4}
\frac{2}{p^2\,(1-\xi\,p^2)^2+2\,i\tilde{\tilde\epsilon} \,(1-\xi\,p^2)-\lambda^2}\,
\end{eqnarray}

\medskip
\noindent
The result of evaluating these integrals  is given
in Appendix, eqs.(\ref{eqII})-(\ref{eqIIend}).
We take into account that $\tilde \epsilon=-\epsilon<0$ and
$\tilde{\tilde\epsilon}=\epsilon>0$ which are important for the UV
behaviour, as they  involve different Wick rotations
 and thus additional minus relative signs.
We provide below the result in the limit when $\xi
\lambda^2\ll 1$  and $\xi\,g^2\,M^2\ll 1$. After  adding
together the contributions above, one has
\medskip
\begin{eqnarray}\label{OR3}
\Delta m_{\varphi_0}^2=\frac{5 \,g^2}{\pi^2} 
\,\, \,(2 g^2\,M^2)^2\,\xi^2\,\Lambda^2+\cO(\ln\Lambda)
\end{eqnarray}

\medskip
\noindent
Similar to the  case of non-vanishing $\xi_1$, we found  that
in the presence of  supersymmetric higher derivative operators, 
 supersymmetry breaking - although spontaneous -  brings in,
 nevertheless,  a quadratic divergence
to the one-loop self energy of the scalar field. 
In the limit of restoring supersymmetry ($g M=0$), the quadratic
divergence is absent, as it should be the case. 
Also, in the special case of decoupling of
the higher derivative operators when their scale is much larger then
the UV scale (i.e. $\xi\Lambda^2\ll 1$) this 
divergence is again absent, as expected. 

In conclusion and  somewhat surprisingly, 
spontaneous supersymmetry breaking in the presence of higher
derivative supersymmetric operators is no longer soft and quadratic
divergences are present, with a  coefficient equal to the ratio of 
the parameter related to the amount of supersymmetry 
breaking  and the scale of higher derivative
operators. The need for a high scale of the higher derivative operators 
discussed earlier, ensures ultimately 
 a small value  of the coefficient of the quadratic divergences.
Moreover, the quadratic divergence has a coefficient which is 
suppressed by the power 4 of the scale $M_*\equiv 1/\sqrt{\xi}$ of the higher derivative
operator.

\subsection{Wess-Zumino model with soft breaking terms and higher derivatives.}\label{WZ}

In this section  we investigate
the extent to which our previous findings for the supersymmetric case
remain true in the case of
{\it explicit} breaking of supersymmetry, as opposed to the
spontaneous breaking analysed.
We consider the Wess-Zumino  model with a  soft breaking
term and
extended  with supersymmetric higher derivative operators. We  
examine the one-loop  self-energy of the scalar field.
 The action   is
\smallskip
\begin{eqnarray}\label{WZaction}
\cL&=&\int d^4\theta \,\Phi^\dagger \Big(1+ \xi \,\Box\Big)\Phi
+\bigg\{\int d^2\theta \bigg(\frac{1}{2} \,m \Phi^2 + \frac{1}{3}\lambda
  \,\Phi^3\bigg)+ c.c.\bigg\}-m^2_{0}\,\varphi\, \varphi^*\nonumber\\[9pt]
&=&
F^* \Big( 1+\xi {\Box} \Big) \,F
- \varphi^*\Box \Big( 1+\xi {\Box} \Big) \,\varphi
+i \partial_\mu \bar \psi  \,\bar\sigma^\mu
\Big( 1+\xi {\Box} \Big)  \psi\nonumber\\[9pt]
&+&\Big( \,
\frac12  m \,\big( 2 \varphi\,F -\psi\psi \big)+\lambda\, 
\big(\varphi^2 \,F-\varphi\,\psi^2\big)
+c.c.\Big)-m^2_{0}\,\varphi\, \varphi^*
\end{eqnarray}

\medskip
\noindent
with $\Phi\equiv(\varphi,\psi,F)$ and a notation similar to that in
the O'Raifeartaigh model.

The last term in (\ref{WZaction}) breaks supersymmetry softly.
 The field $F$  is dynamical  and has  a
ghost field partner ($\Box F$). The off-shell counting of the bosonic and fermionic 
real degrees of freedom works as in the usual Wess-Zumino model, but now each field
has a (ghost) counterpart: $\varphi$ (2), $\Box
\varphi$ (2), $F$ (2), $\Box F$ (2), $\psi$ (4), $\Box \psi$ (4).

The relevant propagators in the presence of higher
derivative term are found as in previous section.
For example one 
writes the quadratic terms in the action
in  basis $\vec \sigma \equiv (\varphi, F^*)$ as $\cL=(1/2)\, 
\vec\sigma\, \cM \,\vec\sigma^{* T}$
and then invert the matrix $\cM$. One finds 
\smallskip
\begin{eqnarray}\label{ss}
<\varphi\,\varphi^*>&=&\frac{-i\,(1+\xi \Box-i\,\tilde\epsilon\,\xi)}{
[\,\Box (1+\xi \Box)+m_0^2-i\epsilon]\,
(1+ \xi \Box -i\,\tilde\epsilon\,\xi)\, +m^2},\nonumber
\\[9pt]
<F\,F^*>&=&\frac{i\,(\Box\,(1+\xi \Box) +m_0^2-i\,\epsilon)}{
[\,\Box (1+\xi \Box)+m_0^2-i\epsilon]\,
(1+ \xi \Box -i\,\tilde\epsilon\,\xi)\, +m^2},
\end{eqnarray}

\medskip
\noindent
with our usual convention that
 $\epsilon>0$. In the above eqs, the $i \epsilon$-prescription is in agreement with that 
for  $\xi=0$ and ultimately read from that of
 the propagator of $\varphi$ alone (with $m=0$),
entering the mass mixing matrix;
 there is also a $i \tilde \epsilon$ prescription, 
which originates from the propagator for the $F$  field in the absence
of any mixing by $m$.  For the  propagator of the field $F$, comparison 
of the higher derivative terms in Wess-Zumino model  against
the second term in (\ref{action_4g}) suggests that  $\tilde
\epsilon =-\epsilon$; let us however keep  the sign of $\tilde \epsilon$
arbitrary, as we did in the O'Raifeartaigh model.
Finally, eqs.(\ref{ss}) are written  in the presence  of
the supersymmetry breaking term, while if supersymmetry is unbroken one simply
sets $m_0=0$. For fermions  one finds the 
  propagator
\medskip
\begin{eqnarray}\label{fermion-propagator}
<\psi^A \overline \psi^{\dot B} > & = & 
(1+\xi\,\Box) \,\sigma^{\mu\,\,A \dot B} \,\, \partial_\mu\,
\frac{1}{\Box (1+\xi \,\Box)^2+m^2-2\,i\,\tilde{\tilde\epsilon}\,
  (1+\xi \,\Box)},
\end{eqnarray}

\medskip
\noindent
 A similar expression exists for 
$<\overline \psi_{\dot A} \psi_{B} >$, but with the matrix 
$\sigma^{\mu\,\,A \dot B}$ 
replaced by $\overline\sigma_{\dot A  B}^{\mu}$.
For the fermionic propagators, the prescription is
$\tilde{\tilde\epsilon}=\epsilon>0$, 
and is found similarly to the O'Raifeartaigh model. 
 In the limit $\xi=0$ one  recovers the usual propagator for Weyl
fermions. 

\begin{figure}[t]{
\centerline{\psfig{figure=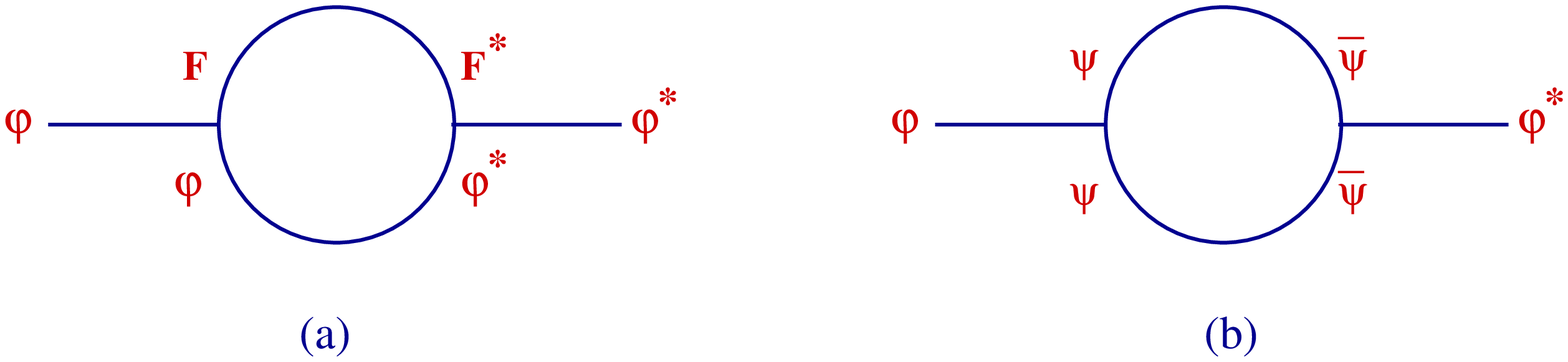,height=0.9in,width=4.8in,angle=0}}
\def\baselinestretch{1.1}
\caption{\small The  one-loop Feynman diagrams (off-shell) which contribute 
to the scalar field self-energy for scalar field $\varphi$, in the
Wess-Zumino model with higher derivative terms.} 
\label{fig1}}
\end{figure}

Using these propagators, one
obtains the following one-loop contributions, see Figure~\ref{fig1},
to the mass of the scalar field $\varphi$, for vanishing external momentum:
\medskip
\begin{eqnarray}
(a)&=& 4 \,\,(i\lambda)^2  \,\,\int \frac{d^d p}{(2\pi)^d}
\frac{-(1-\xi\,p^2-i\,\tilde\epsilon \,\xi)
\,[\,p^2 (1-\xi\,p^2)-m_0^2+i\,\epsilon\,]}{
[\,(\,p^2 (1-\xi\,p^2)- m_0^2+i\,\epsilon )\,\, (1-\xi\,p^2-i\,\tilde\epsilon
\,\xi)-m^2\,]^2}
\\[10pt]
(b)&=& - 2\,\, (-i \lambda)^2 \int \frac{d^d p}{(2\pi)^d}
\frac{(1-\xi\,p^2)^2 \,\,tr (\sigma^\mu \overline \sigma^\nu) (-p_\mu p_\nu)}{
[\,p^2\,(1-\xi \,p^2)^2-m^2+2\,i\,\tilde{\tilde\epsilon}\,(1-\xi\,p^2)]^2},
\end{eqnarray}

\medskip
\noindent
for bosons and fermions respectively, with  $tr(\sigma^\mu\overline
\sigma^\nu)=2\,\eta^{\mu\nu}$. 

For our purposes of investigating the UV behaviour of the self-energy correction,
it is sufficient to consider the  massless case, $m=0$, when the above
loop integrals simplify considerably, without changing their  UV
behaviour. In that case we have\footnote{For (b) we use
$p^2(1-\xi   \,p^2)+2i\tilde{\tilde\epsilon}
=-\xi\,(p^2+2i\,\tilde{\tilde\epsilon})\,(p^2-1/\xi-2i\,\tilde{\tilde\epsilon})+
\cO(\tilde{\tilde\epsilon}^2)$},
\medskip
\begin{eqnarray}\label{WZintegral2}
(a)&=& 4 \,\,(i\lambda)^2  \,\,\int \frac{d^d p}{(2\pi)^d}
\frac{-1}{(-\xi)\,\,
[\,p^2 (1-\xi\,p^2)- m_0^2+i\,\epsilon ]\,
 (p^2-1/\xi+i\,\tilde\epsilon \,\,)},\qquad (\tilde\epsilon=-\epsilon)\nonumber
\\[10pt]
(b)&=&  2\,\, (-i \lambda)^2 \int \frac{d^d p}{(2\pi)^d}
\frac{2}{(-\xi)\,\,
[\,p^2 (1-\xi\,p^2)+2\,i\,\tilde{\tilde\epsilon} ]\,
 (p^2-1/\xi-2\,i\,\tilde{\tilde\epsilon} \,\,)},\qquad \quad 
(\tilde{\tilde\epsilon}=\epsilon)
\end{eqnarray}

\medskip
\noindent
One again observes that if supersymmetry is restored, one must have
that $\tilde\epsilon=-\tilde{\tilde\epsilon}=-\epsilon$, which is
consistent with our choice and as 
already encountered in the O'Raifeartaigh model.

Remembering that (a) and (b) are
contributions to $-i\Delta m_{\varphi}^2$, we find
\medskip
\begin{eqnarray}\label{mmm}
\Delta m^2_{\varphi}&=&\!\!\frac{\lambda^2}{4\pi^2}\,\,
\bigg\{(3\,\xi \,m_0^2)\, 2 \Lambda^2  -
m_0^2\,\,\Big[\ln(1+\xi\,\Lambda^2)-\frac{\sigma}{2}\,(\sigma-4)\Big]
-m_0^2\,(1+4\xi\,m_0^2)\,\ln(1+\Lambda^2/m_0^2)\nonumber\\[12pt]
&&\quad +\,\bigg[\,\frac{m_0^2\,\Lambda^2}{\Lambda^2+m_0^2}-
\frac{m_0^2}{3}\Big(\sigma\,(18-6\,\sigma+\sigma^3)+12\,\ln(1+\Lambda^2\,\xi)\Big)
\bigg] \,\xi\,m_0^2+\cO\Big((\xi\,m_0^2)^2\Big)\bigg\}
\end{eqnarray}

\bigskip
\noindent
where $\sigma=\Lambda^2\xi/(1+\Lambda^2\xi)$.
This result is valid for $\xi\,m_0^2\ll 1$, but for
an arbitrary relation between $\Lambda$ and $\xi$. For the result
without this approximation see the Appendix, eq.(\ref{pp}).

We therefore obtain, similar to the case of the O'Raifeartaigh model, that
a quadratic divergence is present in the overall bosonic and fermionic
contributions, equal to $(3\,\xi \,m_0^2)\, 2 \Lambda^2$. 
All the remaining terms in (\ref{mmm}) arise from expansions (for $\xi
m_0^2\ll 1$) of
logarithms of arguments involving ratios of $\Lambda^2$ and
mass terms function  of $\xi$ and  $m_0^2$. 
The quadratic divergence  is
present despite the soft   nature of supersymmetry breaking,
 and is due to the fact that we considered this  breaking in
the presence of higher derivative  supersymmetric terms.
In the limit of restoring supersymmetry, $m_0\ra 0$ this divergence
and in fact the entire quantum correction is
absent, as expected.  The coefficient of our quadratic divergence is
proportional to the amount of supersymmetry breaking ($m_0$), and inverse
proportional to the scale $M_*^2=1/\xi$ of the higher derivative operator.
These results  remain valid in the case of a  non-zero
mass shift $m$ for  fermions/bosons.

It is worth noticing that, unlike the case of O'Raifeartaigh model, in
the present case the coefficient of the quadratic divergence is
less suppressed, by $M_*^2$ ($M^2_*=1/\xi$), rather than $M_*^4$. The
origin of this difference can be traced back to the presence in the
 Wess Zumino-model
of the soft breaking term $m_0^2 \,\varphi\,\varphi^*$, while in the
spontaneous breaking, its counterpart involved a bilinear term 
$g^2 M^2 \varphi_0^2+c.c.$ rather than  $g^2 M^2 \varphi_0 \varphi_0^*$,
 see (\ref{newL}). 

Let us finally take the special limit  $\xi\,\Lambda^2\ll 1$, which 
decouples the higher derivative operators in the classical action.
In this case  only a logarithmic correction (in the UV cutoff)
 survive in (\ref{mmm}) of type $-m_0^2\ln(1+\Lambda^2/m_0^2)$,
 the higher derivative operator and associated  ghost fields
 contributions decouple, to recover  that no quadratic divergences
 are present~\cite{GG} in
 softly broken supersymmetry in an action without higher dimensional terms.

We conclude this section with a remark which applies to both
O'Raifeartaigh and Wess-Zumino models.  Our analysis is somewhat
restrictive in that  we considered only the role  of one particular 
 higher dimension
(derivative) operator on the UV behaviour of the scalar field;
however,  other operators of same or  lower mass
dimension can be present and should   be included for a
 complete study, with possibly different conclusions.
For example one can have additional
 operators such as
$\int d^2\theta\Phi\Box^2 \Phi$ or $\int d^4\theta 
(\Phi^\dagger \Phi)^2$, suppressed by additional powers of a 
mass scale. Such new scale introduces additional parameters in the theory,
unless this is taken equal to $M_*$.
 Such terms can change significantly our one-loop results
and bring in additional, interesting effects.

\section{Final Remarks and Conclusions}

In this work we discussed the role that higher derivative operators 
 play at the  quantum level, and their implications, for both
non-supersymmetric and supersymmetric 4D theories. Such operators are in
general expected  in  effective theories, and in models of
 compactification and this motivated this study, despite  the problems
that theories with higher derivative operators may have.

We first considered a 4D scalar field theory with  higher
derivatives, and stressed the important role of  a  well-defined
partition function in the Minkowski space at {\it all momentum scales},
 for loop calculations and for
 analytical continuation to the Euclidean space to exist and 
be unambiguously defined.
The  loop corrections to the self-energy of the scalar field were
computed in the presence of higher derivative terms, to show that
these do not necessarily improve the UV behaviour of the theory, as
usually considered. This is because Minkowski space-time 
power-counting criteria for convergence do not always remain true
 in higher order theories.
The reason for this is the  relation Minkowski-Euclidean analytical
continuation,  which involves Wick
rotations in opposite senses, such that ultimately ghost-like and
particle-like degrees of freedom contributions add up  (rather than cancel)
to the UV behaviour of the scalar field self-energy. 

We also discussed the relation between a Minkowski and an Euclidean theory
both with higher derivative terms, and similar action at the tree
level, to show that their relation is complicated at loop level by
the presence of higher derivative terms. This complication arises due
to  the additional poles that  higher derivative terms  induce in the Minkowski
spacetime. As a result of their presence, 
 the two theories can give similar
result for the one-loop self energy, although they do not have an
identical set of operators or these can come with different
coefficients.
It also turns out that not even the Euclidean theory with higher 
derivatives is finite or only logarithmic divergent. Instead,
 it can also have quadratic divergences
in the case that other dimension six-operators are included, in
addition to the higher derivative kinetic term. 

 The analytical  continuation of our loop corrections 
derived in the Minkowski space-time
 is further affected at the dynamical level due to a field-dependence of
 the poles of the  Green functions of the particle-like states,
for curvatures of the  potential of order unity in ghost mass units.
 The one-loop scalar potential in $\lambda \phi^4$ theory in the presence of a
single  higher derivative  operator is shown to have infinitely many
 counterterms, while for a large mass  of the ghost the usual
4D renormalisation is recovered. There exists an interesting
 cancellation of the quartic
divergence in the (field dependent part of the) 
one-loop potential, between the scalar field and its ghost counterpart 
contributions, and this suggests that the higher derivative operators
 may play a role for the cosmological constant problem.
The cancellation is a property specific to the Minkowski  space-time 
formulation of the theory and its partition function 
convergence, and is absent in a counterpart
Euclidean theory with higher derivatives.

Our study also considered supersymmetric models with higher derivative
terms in the action. In the case of O'Raifeartaigh models with
spontaneous supersymmetry breaking and (supersymmetric)
higher derivative terms it was shown that, despite the soft nature of
the breaking, quadratic divergences are nevertheless present. 
The coefficient of this divergence is given by the ratio of the 
amount of supersymmetry breaking  to the scale of higher derivative
operators, and thus vanishes when restoring supersymmetry or when
decoupling the higher derivative terms.
Similar results hold true in the case of Wess-Zumino model
with higher derivative terms and soft (in the traditional sense), 
explicit supersymmetry breaking
terms and this was investigated in detail.
The emergence of these quadratic divergences is related to the
presence of ghost fields in the loop corrections; these corrections
vanish when such fields acquire a  mass much larger than the UV cutoff
scale (the decoupling limit) and then softly broken supersymmetry is restored.

Although our work considered toy-models only,
it may be  interesting to think of the implications of these findings
for phenomenology and for the hierarchy problem in realistic models.
 Despite many  problems that theories with
higher derivative terms may have conceptually or phenomenologically
(their stability, possible 
unitarity violation, etc) let us  compare the leading correction 
$\xi\, m_0^2 \,\Lambda^2$
 found for the scalar field self-energy in Wess-Zumino model,
 against its counterpart in  the supersymmetric versions
 of the Standard Model, equal to $m_0^2 \ln\Lambda^2/m_0^2$, (we assume
 similar values for the soft mass $m_0$). The two corrections would be of
 similar order of magnitude provided that $\xi \Lambda^2\sim
 \cO(10-100)$. This would  set the scale of the higher derivative
 operator within a factor of 10 or so below the cutoff scale/Planck scale. Such scale
 can be high enough to remove any conceptual problems that higher
 derivative operators might bring. In the case of O'Raifeartaigh
 models of supersymmetry breaking, comparing the one-loop correction 
to the scalar field mass,  which is $(\xi\, m_0^2)^2 \,\Lambda^2$
to $m_0^2 \ln\Lambda^2/m_0^2$,  gives  $\xi\,\Lambda \sim
\cO(10^{-5}-10^{-4})\,GeV^{-1}$ after assuming a TeV-scale value for
$m_0$. As a result, in this case the scale of higher derivative
operators can be significantly lower, in the range of
 intermediate energies, $\cO(10^{11}-10^{12})$ GeV.

The results obtained in this work may be extended to higher dimensional theories
with various supersymmetry breaking mechanisms, in the presence of
higher derivative operators on the ``visible'' brane or in the
bulk.  This is interesting since such
operators can be generated dynamically during compactification, thus
their effects need to be taken into account.
It is likely that the supersymmetry breaking effects seen here
are present in such theories too, with potentially interesting
implications  for theory and  phenomenology.

\vspace{0.7cm}
\noindent
{\bf Acknowledgements:  }
D.G. thanks Hyun Min Lee (DESY) for many interesting and useful
discussions on higher derivative theories and research
collaboration on related topics. D.G. acknowledges the financial 
support from the  RTN European Program MRTN-CT-2004-503369
 ``The Quest for Unification'' to attend  the ``Planck 2006''
conference  (Paris, May 2006) where part of this work was completed.
This work was supported in part by the European Commission under the
RTN contracts MRTN-CT-2004-503369, MRTN-CT-2004-005104,
the European Union Excellence Grant, MEXT-CT-2003-509661,
CNRS PICS no. 2530 and 3059 and in part by the INTAS contract 03-51-6346.

\section*{Appendix}
 
\def\theequation{\thesubsection-\arabic{equation}}
\def\thesubsection{A}
\setcounter{equation}{0}
\label{appendixA}
 
{\bf I.   } We have the following derivatives  for the scalar potential
obtained in the text, eq.(\ref{final-one-loop})
\begin{eqnarray}\label{eqI}
\frac{\partial^2 \Delta U_2(\phi)}{\partial\phi^2}\bigg\vert_{\phi=0}
&=&
\frac{-\lambda\, c_o}{2 \xi (1-4\, \xi\, m^2)^{1/2}}
\bigg[
2 \,\xi \,
m_-^2 \ln\frac{m_-^2}{\mu^2}
+2 \,\xi\, m^2_+ \ln\frac{m^2_+}{\mu^2}+1\bigg]\nonumber\\[12pt]
\frac{\partial^4 \Delta U_2(\phi)}{\partial\phi^4}
\bigg\vert_{\phi=0}
&=&\frac{-3 \,\lambda^2 \, c_o}{(1-4\,\xi\,m^2)^{3/2}}
\bigg[1+ \,\ln \frac{m^2}{\xi \,\mu^4}\bigg]
\end{eqnarray}

\vspace{0.5cm}
\noindent
{\bf II.} The propagators used in the case of the O'Raifeartaigh
model, eqs.(\ref{wrwr}),  are given below. For $\lambda=0$, $M=0$ one
recovers the propagators in the absence of mass mixing terms, given in 
eqs.(\ref{aa1}) to (\ref{fermionpropagators}), together with their
prescriptions for the poles. The results are:

\smallskip
\begin{eqnarray}\label{prop1}
<F_0\,F_0^*>=i\,\frac{\Box(1+\xi_2\,\Box)-i\epsilon}{
\Big(\Box(1+\xi_2\,\Box)-i\,\epsilon\Big)\Big(
1+\xi_0\Box-i\,\tilde\epsilon\,\xi_0\Big)+\lambda^2}
\end{eqnarray}
and 
\begin{eqnarray}\label{prop2}
<\varphi_0\,\varphi_0^*>=\frac{-i\, (1+\xi_2\,(\Box-i\,\tilde \epsilon\,))\,
\,[\lambda^2 + (1+\xi_2 \,(\Box-i\,\tilde \epsilon))\,\,(\Box
  (1+\xi_0\Box)-i\,\epsilon)\,]}{
[\lambda^2 + (1+\xi_2 \,(\Box-i\,\tilde \epsilon))\,\,(\Box
  (1+\xi_0\,\Box)-i\,\epsilon)\,]^2- (2 g^2 M^2)^2 \,(1+\xi_2\,(\Box -i\,\tilde
\epsilon))^2}
\end{eqnarray}

\medskip
\noindent
Finally
\begin{eqnarray}\label{prop3}
<\overline \psi_0\psi_0>= \frac{ \partial\!\!\slash \,[
    \Box\,(1+\xi_2\,\Box)-i\,\tilde{\tilde \epsilon}\,]}{
(\Box\,(1+\xi_0\,\Box)-i\,\tilde{\tilde \epsilon}\,)\,[\,\Box\,(1+\xi_2
\,\Box)-i\,\tilde{\tilde\epsilon}\,] +\lambda^2\,\Box}
\end{eqnarray}

\medskip
\noindent
where $\not\!\!\partial\equiv \partial^{\mu}\sigma_\mu$.
Above we used  $<F_0 F_0^*>=(\cN^{-1})_{44}$ 
where $\cN^{-1}$ is the inverse of matrix $\cN$ which can be read from
 the Lagrangian of eqs.(\ref{O'Rmodel}), (\ref{newL}): 
$\cL\supset (1/2)\, \,\vec w \,\cN\,{\vec w}^{*\, T}$ with $w$  the vector 
basis $w\equiv (\varphi_2^*, \varphi_2, F_0^*, F_0)$.
Further,  $<\varphi_0\,\varphi_0^*>=(\cM^{-1})_{22}$, where the matrix $\cM$ is read
 from   $\cL\supset (1/2)\,\, \vec\gamma\,
 \cM\,{\vec \gamma}^{*\, T}$, with $\vec\gamma\equiv (\varphi_0^*,
 \varphi_0, F_2^*, F_2)$. Finally
 $<\overline\psi_0\,\psi_0>=(\cP^{-1})_{11}$, 
 where $\cP$ can be read
from $\cL\supset  (1/2)\,\,\vec\delta \,\cP\, {\vec \delta}^{*\, T}$
 with $\vec \delta \equiv (\overline \psi_0,\psi_0, \overline \psi_2,
 \psi_2)$. The poles prescriptions are dictated by those in the
 absence of any mass mixing terms in the Lagrangian, and are discussed
 in the main text after eq.(\ref{fermionpropagators}).

\vspace{0.5cm}
\noindent
{\bf III.} The integrals encountered in the text,
eqs.(\ref{OR2}), can be written 
\medskip
\begin{eqnarray}\label{eqII}
\cI=\int \frac{d^d p}{(2\pi)^d}
\frac{1}{(p^2-m_1^2)\,(p^2-m_2^2)\,(p^2-m_3^2)}
=\sum_{i=1}^{3}  \frac{1}{\delta_{ij}\delta_{ik}}
\int\frac{d^d p}{(2\pi)^d}\frac{1}{p^2-m_i^2}
 \quad j\not=k\not=i\not=j\quad
\end{eqnarray}

\medskip
\noindent
where   $m_i^2$ include the imaginary part of the propagator
($\epsilon$ prescription), needed for  Wick rotations
(clock-wise/anti-clockwise, depending on the sign). We denote the
sign of this imaginary part of  $m_i^2$ by $u_i$  and also introduce
  $\delta_{ij}=m_i^2-m_j^2$.
In the cutoff regularisation, $d=4$, one has
\begin{eqnarray}
\cI=\frac{i\pi^2}{(2\pi)^4}
\sum_{i=1}^{3}  \frac{u_i}{\delta_{ij}\delta_{ik}} \,f(\Lambda,m_i);
\quad j\not=k\not=i\not=j;\qquad
f(\Lambda^2,m_i^2)=\Lambda^2-m_i^2\ln (1+\Lambda^2/m_i^2)
\end{eqnarray}

\medskip
\noindent
For integral (b) of eqs.(\ref{OR2}) we thus obtain
\medskip
\begin{eqnarray}\label{a8}
(b)&=&(2ig)^2 \frac{1}{2}\int\frac{d^4 p}{(2\pi)^4}
\frac{-1}{(1-\xi p^2+i \epsilon\,\xi)\,[p^2\,(1-\xi\,p^2)+\rho+i\epsilon]-\lambda^2}
\bigg\vert_{\rho=2g^2 M^2}+(\rho\ra -2 g^2 M^2)
\nonumber\\[12pt]
&=&
\frac{i\,g^2}{4\pi^2\xi^2}\frac{1}{2}
\bigg\{\frac{-f(\Lambda^2,m_3^2)}{\delta_{31}\delta_{32}}\,
+
\frac{f(\Lambda^2,m_1^2)}{\delta_{31}\delta_{21}}\,-
\frac{f(\Lambda^2,m_2^2)}{\delta_{32}\delta_{21}}\,\bigg\}_{\rho=2g^2M^2}
+(\rho\ra -2 g^2 M^2)
\end{eqnarray}

\medskip
\noindent
where $m_{1,2,3}$ are given by the roots of 
\begin{eqnarray}\label{a9}
&&(1-\xi p^2+i
 \epsilon\,\xi)\,[p^2\,(1-\xi\,p^2)+\rho+i\epsilon]-\lambda^2=
\xi^2\,(p^2-m_1^2)(p^2-m_2^2)(p^2-m_3^2),
\end{eqnarray}

\medskip
\noindent
Notice that the lhs accounts for one particle-like propagator
involving $p^2 +i \epsilon$,  and two
ghost-like ones, which depend on $p^2-i\epsilon$.
In the approximation $\xi\lambda^2\ll 1,\,\,\xi\,\rho^2\ll 1$  
the roots $m_{1,2,3}$ are
\smallskip
\begin{eqnarray}\label{m123}
m_{1,2}^2&=&\frac{1}{\xi}(1\pm \sqrt{\xi\,\lambda^2})
+\frac{1}{2} \,(\rho  -\lambda^2)
+i\epsilon
\mp
\frac{\xi^{\frac{1}{2}}}{8\sqrt{\lambda^2}}\,(6\lambda^2\rho-\rho^2-5\lambda^4)+
\frac{\xi}{2} \,(3\lambda^2\rho-\rho^2-2\lambda^4)\nonumber\\[10pt]
&\mp&\frac{\xi^{\frac{3}{2}}
\sqrt{\lambda^2}}{128\lambda^4}\,(-231\lambda^8+420\lambda^6\rho+\rho^4
-210\lambda^4\rho^2+20\lambda^2\rho^3)+\cO(\xi^{2})
;\nonumber\\[12pt]
m_3^2&=&\lambda^2-\rho-i\,\epsilon+\xi\,(2\lambda^4-3\lambda^2\rho+\rho^2)+
\cO(\xi^{3/2}),
\end{eqnarray}
where upper (lower) signs correspond to $m_1$ ($m_2$) respectively.
From  $m_{1,2}$ above we find  $u_3=-1$, $u_1=u_2=1$,  used in
eq.(\ref{a8}), in agreement with our earlier observation of one
particle-like and two ghost-like roots/propagators.

Further, integral (a) of  eqs.(\ref{OR2}) is  given by (\ref{a8}) 
with $\rho=0$ in both terms. Integral (c) of  (\ref{OR2}) is equal to
$(-2)$ times the result (a).
Adding together  (a), (b), (c) of (\ref{OR2}) we obtain, with
$(a)+(b)+(c)=-i \Delta m^2_{\varphi_0}$ that
\medskip
\begin{eqnarray}\label{z2}
\Delta m_{\varphi_0}^2&=&
\frac{-g^2}{4\pi^2\xi^2}\frac{1}{2}
\bigg\{\bigg[\frac{-f(\Lambda^2,m_3^2)}{\delta_{31}\delta_{32}}\,+
\frac{f(\Lambda^2,m_1^2)}{\delta_{31}\delta_{21}}\,-
\frac{f(\Lambda^2,m_2^2)}{\delta_{32}\delta_{21}}\,\bigg]
\bigg\vert_{\rho=2g^2M^2}\!\!\!\!+(\rho\ra -2 g^2 M^2) \bigg\}\nonumber\\[9pt]
&-& \Big\{\rho\ra 0\Big\}
\end{eqnarray}

\medskip
\noindent
 The exact coefficient of the  quadratic divergence can then be read from
\medskip
\begin{eqnarray}\label{z1}
\Delta m_{\varphi_0}^2=
\frac{-g^2}{4\pi^2\,\xi^2}\,
\bigg\{
\frac{1}{\delta_{31}\delta_{32}}\bigg\vert_{\rho=0}-\frac{1}{2}
\frac{1}{\delta_{31}\delta_{32}}\bigg\vert_{\rho=2 g^2 M^2}
-\frac{1}{2}
\frac{1}{\delta_{31}\delta_{32}}\bigg\vert_{\rho=-2 g^2 M^2}
\bigg\}\,(2\,\Lambda^2)+\cO(\ln\Lambda)\qquad
\end{eqnarray}

\medskip
\noindent
The last two eqs  can be simplified further in 
 the limit $\xi\,\lambda^2\ll 1$,  $\xi\, g^2\,M^2\ll 1$, by using 
(\ref{m123}) up to $\cO(\xi^{3/2})$ terms in $\delta_{31}$,
 $\delta_{32}$;
 in this approximation the UV  quadratic term has a coefficient 
\medskip
\begin{eqnarray}\label{eqIIend}
\Delta m_{\varphi_0}^2&=&
\frac{g^2}{4\pi^2}\, 20 \,\xi^2 \,(2 g^2 M^2)^2\,\Lambda^2
+\cO(\ln\Lambda)
\end{eqnarray}
The quadratic term  received contributions from
both values $\rho=\pm 2 g^2 M^2$, each contributing to half of its  coefficient.
This results was used in the text, eq.(\ref{OR3}).

\bigskip
\noindent
{\bf IV.   } Integral (a) in (\ref{WZintegral2}) is (here one can set $v=\pm 1$):

\begin{eqnarray}\label{pp}
(a)&=& 4 \,\,(i\lambda)^2  \,\,\int \frac{d^d p}{(2\pi)^d}
\frac{-1}{(-\xi)\,\,
[\,p^2 (1-\xi\,p^2)- m_0^2+i\,\epsilon ]\,
 (p^2-1/\xi+i\,\tilde\epsilon \,\,)},\qquad
(\tilde\epsilon=v\,\epsilon, \,\,v=\pm 1)
\nonumber\\[12pt]
&=&
\frac{-i \lambda^2}{4\pi^2}
\frac{1}{\xi^2\,(M_+^2-M_-^2)}
\,\bigg\{\,\,
\Lambda^2\,\bigg[ \frac{1-v}{M_+^2}+ \frac{1+v}{M_-^2}\bigg]
+\frac{v}{\xi}\bigg[\frac{1}{M_+^2}-\frac{1}{M_-^2}\bigg]
\,\ln\Big(1+\Lambda^2\,\xi\Big)\nonumber\\[12pt]
&&\qquad\qquad\qquad\qquad\quad
-\frac{M_-^2}{M_+^2}\,\ln\Big(1+\frac{\Lambda^2}{M_-^2}\Big)
-\frac{M_+^2}{M_-^2}\,\ln\Big(1+\frac{\Lambda^2}{M_+^2}\Big)\bigg\}
\end{eqnarray}
where
\begin{eqnarray}
M_\pm^2=\frac{1}{2\xi}\,\Big(1\pm \sqrt{1-4\,\xi\, m_0^2}\,\Big)
\end{eqnarray}
The exact coefficient of the  quadratic term is  read from the above
eq with $v=-1$
\begin{eqnarray}
(a)=\frac{-i \lambda^2}{4\pi^2}
\frac{4}{1+\sqrt{1-4 \xi \,m_0^2}}
\frac{1}{\sqrt{1-4\xi\,m_0^2}}\,\,\Lambda^2 +\cO(\ln\Lambda)
\end{eqnarray}

The result quoted in (\ref{mmm}) is derived by using the one
above, for $v=-1$, $\xi\,m_0^2\ll 1$, but for an arbitrary value of
 $(\xi \,\Lambda^2)$.  In that case one obtains
\medskip
\begin{eqnarray}\label{ppp}
\!\!\!(a)&\!=\!&\!\!\frac{-i \lambda^2}{4\pi^2}\,\,
\bigg\{(1+3\,\xi \,m_0^2)\, 2 \Lambda^2  -
m_0^2\,\,\Big[\ln(1+\xi\,\Lambda^2)-\frac{\sigma}{2}\,(\sigma-4)\Big]
-m_0^2\,(1+4\xi\,m_0^2)\,\ln(1+\frac{\Lambda^2}{m_0^2})\nonumber\\[12pt]
&-&\!\!\frac{\sigma}{\xi}\,+\bigg[\,\frac{m_0^2\,\Lambda^2}{\Lambda^2+m_0^2}-
\frac{m_0^2}{3}\Big(\sigma\,(18-6\,\sigma+\sigma^3)+12\,\ln(1+\Lambda^2\,\xi)\Big)
\bigg] \,\xi\,m_0^2+\cO\Big(\xi\,m_0^2)^2\Big)\bigg\}
\end{eqnarray}

\medskip
\noindent
where $\sigma=\Lambda^2\xi/(1+\Lambda^2\xi)$.
To evaluate integral (b) in (\ref{WZintegral2}), one takes
$m_0\!\ra\! 0$ on the above results, and multiplies them by~-1. By
adding 
the fermionic counterpart  to (\ref{ppp}), one obtains the result in (\ref{mmm}).


\begin{thebibliography}{111}  

\bibitem{swh}
S. W. Hawking, in {\it ``Quantum Field theory and Quantum Statistics:
Essays in Honour of the 60 th Birthday of E.S. Fradkin''},
eds. A.Batalin, C.J.Isham, C.A. Vilkovisky, Hilger, Bristol, UK
(1987).

\bibitem{Hawking:2001yt}
S.~W.~Hawking and T.~Hertog,
{\it ``Living with ghosts,''}
Phys.\ Rev.\ D {\bf 65} (2002) 103515
[hep-th/0107088].

\bibitem{Simon:1990ic}
J.~Z.~Simon,
{\it ``Higher Derivative Lagrangians, Nonlocality, Problems And Solutions,''}
Phys.\ Rev.\ D {\bf 41} (1990) 3720.

\bibitem{Smilga:2005gb}
A.~V.~Smilga,
{\it ``Ghost-free higher-derivative theory,''}
hep-th/0503213;
{\it ``Benign vs. malicious ghosts in higher-derivative theories,''}
Nucl.\ Phys.\ B {\bf 706} (2005) 598
[hep-th/0407231].

\bibitem{deUrries:1998bi}
F.~J.~de Urries and J.~Julve,
{\it ``Ostrogradski formalism for higher-derivative scalar field theories,''}
J.\ Phys.\ A {\bf 31} (1998) 6949
[hep-th/9802115];
%
J. Barcelos-Neto, C.P. Natividade, {\it ``Hamiltonian path integral
  formalism with higher derivatives''}, Z. Phys.C 51, 313, 1991.

\bibitem{c3}
P.~D.~Mannheim, A.~Davidson,
{\it ``Dirac quantization of the Pais-Uhlenbeck fourth order oscillator,''}
hep-th/0408104;
{\it ``Fourth order theories without ghosts,''}
hep-th/0001115;

\bibitem{Gosselin:2002fy}
  P.~Gosselin and H.~Mohrbach,
  {\it ``Renormalization of higher derivative scalar theory,''}
  Eur.\ Phys.\ J.\ directC {\bf 4} (2002) 10.

\bibitem{Rivelles:2003jd}
  V.~O.~Rivelles,
  {\it ``Triviality of higher derivative theories,''}
  Phys.\ Lett.\ B {\bf 577} (2003) 137
  [arXiv:hep-th/0304073].

\bibitem{Pivovarov:2005kh}
  G.~B.~Pivovarov,
  {\it ``Asymptotically free phi**4 in 3+1,''}
  arXiv:hep-th/0510257.


\bibitem{c2}
T. Hamazaki, T. Kugo, {\it ``Defining the Nambu-Jona-Lasinio Model by
  higher derivative kinetic term''}
E-print: arXiv:hep-ph/9405375.
E. Villase\~nor, {\it ``Higher Derivative Fermionic theories''}
E-print: arXiv:hep-th/0203197.
 
\bibitem{Ghilencea:2004sq}
D.~M.~Ghilencea,
{\it  ``Higher derivative operators as loop counterterms in one-dimensional
 field theory orbifolds''}
JHEP {\bf 0503} (2005) 009 [arXiv:hep-ph/0409214]
  D.~M.~Ghilencea and H.~M.~Lee,
  {\it ``Higher derivative operators from transmission
  of supersymmetry breaking  on
  S(1)/Z(2),''}
  JHEP {\bf 0509} (2005) 024  [hep-ph/0505187];
  {\it ``Higher derivative operators from Scherk-Schwarz supersymmetry breaking on
  T**2/Z(2),''}
  JHEP {\bf 0512} (2005) 039  [arXiv:hep-ph/0508221];

 \bibitem{nibbelink} S.~Groot Nibbelink and M.~Hillenbach,
  {\it ``Renormalization of supersymmetric gauge theories on orbifolds: Brane gauge
  couplings and higher derivative operators,''}
  Phys.\ Lett.\ B {\bf 616} (2005) 125
  [arXiv:hep-th/0503153];
 S.~Groot Nibbelink and M.~Hillenbach,
  {\it ``Quantum Corrections to Non-Abelian SUSY Theories on Orbifolds,''}
 E-print: arXiv:hep-th/0602155.
                                                                      
\bibitem{Ghilencea:2006qm}
 D.~M.~Ghilencea, Hyun Min Lee, K. Schmidt-Hoberg
{\it ''Higher Derivatives and brane-localised kinetic terms in gauge
  theories on  orbifolds.''}, E-print: arXiv:hep-ph/0604215. 
  D.~M.~Ghilencea,
  {\it ``Compact dimensions and their radiative mixing,''}
  Phys.\ Rev.\ D {\bf 70} (2004) 045018
  [arXiv:hep-ph/0311264].

\bibitem{santamaria}
J.~F.~Oliver, J.~Papavassiliou and A.~Santamaria,
{\it ``Can power corrections be reliably computed in models with extra  dimensions?,''}
Phys.\ Rev.\ D {\bf 67} (2003) 125004  [arXiv:hep-ph/0302083].

\bibitem{Alvarez:2006we}
  E.~Alvarez and A.~F.~Faedo,
  {\it ``Renormalized Kaluza-Klein theories,''}
  arXiv:hep-th/0602150.

\bibitem{RS}
A.\,Lewandowski, R.\,Sundrum,
{\it ``RS1, Higher Derivatives and Stability''}
E-print: arXiv:hep-th/0108025.


\bibitem{Anisimov:2005ne}
  A.~Anisimov, E.~Babichev and A.~Vikman,
  {\it ``B-inflation,''}
  JCAP {\bf 0506} (2005) 006
  [arXiv:astro-ph/0504560].


\bibitem{Gibbons:2003yj}
  G.~W.~Gibbons,
  {\it ``Phantom matter and the cosmological constant,''}
  arXiv:hep-th/0302199.

\bibitem{BMP}
  V.~Branchina, H.~Mohrbach, J.~Polonyi,
  {\it ``The antiferromagnetic Phi**4 model. I: The mean-field solution,''}
  Phys.\ Rev.\ D {\bf 60} (1999) 045006
  [hep-th/9612110]; 
  {\it ``The antiferromagnetic phi**4 model. II: The one-loop renormalization,''}
  Phys.\ Rev.\ D {\bf 60} (1999) 045007
  [hep-th/9612111].

\bibitem{c1}
A.A.Slavnov, {\it ``Renormalisable electroweak model without
  fundamental scalar mesons''}, E-print: arXiv:hep-th/0601125;
  {\it ``Higgs mechanism as a collective effect due to extra dimension,''}
  arXiv:hep-th/0604052.

\bibitem{c5}
K.~Jansen, J.~Kuti and C.~Liu,
{\it ``The Higgs model with a complex ghost pair,''}
Phys.\ Lett.\ B {\bf 309} (1993) 119
[arXiv:hep-lat/9305003].

\bibitem{Stelle}
K.S. Stelle, {\it ``Renormalisation of Higher Derivative Quantum Gravity''}
  Phys. Rev. {\bf D 16} (1977) 953.

\bibitem{Myers:1987yn}
  R.~C.~Myers,
  {\it ``Higher Derivative Gravity, Surface terms and String Theory,''}
  Phys.\ Rev.\ D {\bf 36} (1987) 392.

\bibitem{Ferrara}
S.~Ferrara, B.~Zumino, {\it ``Structure of linearised supergravity and
  conformal supergravity''}, Nucl.\,Phys.\, B134 (1978), 301.


\bibitem{KKP}
N.V.\,Krasnikov, A.B.\,Kyiatkin, E.R.\,Poppitz,
{\it ``Structure of the effective potential in supersymmetric theories
  with higher order derivatives coupled to supergravity''}
Phys.\,Lett.\,B\, 222, (1989) 66.

\bibitem{Antoniadis}
I.~Antoniadis, E.T.\,Tomboulis,  {\it ``Gauge invariance and Unitarity
  in Higher Derivative Quantum Gravity''}, Phys.\,Rev.\,D33, (1986) 2756.

\bibitem{Tomboulis:1983sw}
  E.~T.~Tomboulis,
  {\it ``Unitarity In Higher Derivative Quantum Gravity,''}
  Phys.\ Rev.\ Lett.\  {\bf 52} (1984) 1173.

\bibitem{Hamber:1984kx}
  H.~W.~Hamber and R.~M.~Williams,
  {\it ``Higher Derivative Quantum Gravity On A Simplicial Lattice,''}
  Nucl.\ Phys.\ B {\bf 248} (1984) 392
  [Erratum-ibid.\ B {\bf 260} (1985) 747].

\bibitem{Nojiri:2000gv}
  S.~Nojiri and S.~D.~Odintsov,
  {\it  ``Brane-world cosmology in higher derivative gravity or warped
  compactification in the next-to-leading order of AdS/CFT  correspondence,''}
  JHEP {\bf 0007} (2000) 049
  [arXiv:hep-th/0006232].

\bibitem{Dubovsky:2004sg}
  S.~L.~Dubovsky,
  {\it ``Phases of massive gravity,''}
  JHEP {\bf 0410} (2004) 076
  [arXiv:hep-th/0409124].

\bibitem{Avramidi:1985ki}
  I.~G.~Avramidi and A.~O.~Barvinsky,
  {\it ``Asymptotic Freedom In Higher Derivative Quantum Gravity,''}
  Phys.\ Lett.\ B {\bf 159} (1985) 269.

\bibitem{Eliezer:1989cr}
  D.~A.~Eliezer and R.~P.~Woodard,
  {\it ``The Problem Of Nonlocality In String Theory,''}
  Nucl.\ Phys.\ B {\bf 325} (1989) 389.

\bibitem{Polyakov:1986cs}
  A.~M.~Polyakov,
  {\it ``Fine Structure of Strings,''}
  Nucl.\ Phys.\ B {\bf 268} (1986) 406.

\bibitem{Slavnov:1977zf}
  A.~A.~Slavnov,
  {\it The Pauli-Villars Regularization For Nonabelian Gauge Theories. (In
  Russian),''}
  Teor.\ Mat.\ Fiz.\  {\bf 33} (1977) 210.
  {\it ``Invariant regularization of gauge theories,''}
  Teor.\ Mat.\ Fiz.\  {\bf 13} (1972) 174;
  C.~P.~Martin and F.~Ruiz Ruiz,
  {\it ``Higher Covariant Derivative Pauli-Villars Regularization Does Not Lead To A
  Consistent QCD,''}
  Nucl.\ Phys.\ B {\bf 436} (1995) 545
  [arXiv:hep-th/9410223].
%
  M.~Asorey and F.~Falceto,
  {\it ``On the consistency of the regularization of gauge theories by high
  covariant derivatives,''}
  Phys.\ Rev.\ D {\bf 54} (1996) 5290
  [arXiv:hep-th/9502025];
%
  T.~D.~Bakeyev and A.~A.~Slavnov,
  {\it ``Higher covariant derivative regularization revisited,''}
  Mod.\ Phys.\ Lett.\ A {\bf 11} (1996) 1539
  [arXiv:hep-th/9601092].

\bibitem{Weinberg:1959nj}
  S.~Weinberg,
  {\it ``High-energy behavior in quantum field theory,''}
  Phys.\ Rev.\  {\bf 118} (1960) 838.


\bibitem{Kaplan:2005rr}
  D.~E.~Kaplan and R.~Sundrum,
  {\it ``A symmetry for the cosmological constant,''}
  arXiv:hep-th/0505265.

\bibitem{GG}
L.~Girardello, M.T.~Grisaru, 
{\it ``Soft breaking of Supersymmetry''},   Nucl.\,Phys.\,B194
(1982)~65
\end{thebibliography}
\end{document}